\documentclass[showpacs,aps,prd,nofootinbib,showkeys,superscriptaddress,twocolumn]{revtex4-1}
\usepackage{graphicx}
\usepackage{bm}
\usepackage{amssymb}
\usepackage{amsmath,latexsym}
\usepackage[usenames]{color}
\usepackage{subfigure}
\usepackage{subfigure}
\usepackage{physymb}
\usepackage{slashed}
\usepackage{multirow,array}
\usepackage{mathtools}
\usepackage{mathrsfs}
\usepackage[colorlinks=false,linktocpage=true]{hyperref}
\usepackage{hyperref}
\usepackage[utf8]{inputenc}
\usepackage{lipsum}
\newcommand{\be}{\begin{equation}}
\newcommand{\ee}{\end{equation}}
\newcommand{\bea}{\begin{eqnarray}}
\newcommand{\eea}{\end{eqnarray}}
\newcommand{\beal}{\begin{align}}
\newcommand{\enal}{\end{align}}
\newcommand{\bs}{\begin{subequations}}
\newcommand{\es}{\end{subequations}}
\newcommand{\besp}{\begin{split}}
\newcommand{\eesp}{\end{split}}

\newcommand{\g}{\hat{g}_{\mu\nu}}

\newcommand{\Du}{\hat{\Delta}^{\mu\nu}}

\newcommand{\gub}{SO(3)_q\otimes SO(1,1)\otimes Z_2}

\newcommand{\um}{\hat{u}^\mu}
\newcommand{\uum}{\hat{u}_\mu}
\newcommand{\un}{\hat{u}^\nu}
\newcommand{\unn}{\hat{u}_\nu}
\newcommand{\tmn}{\hat{T}^{\mu\nu}}

\newcommand{\piu}{\hat{\pi}^{\mu\nu}}
\newcommand{\parho}{\partial_\rho}
\newcommand{\pit}{\hat{\pi}}

\newcommand{\pp}{{\hat{p}}}
\newcommand{\po}{\hat{p}_\Omega^2}

\newcommand{\pvv}{\hat{p}_\eta}
\newcommand{\pth}{\hat{p}_\theta}
\newcommand{\pph}{\hat{p}_\phi}
\newcommand{\ph}{\hat{p}^\rho}

\newcommand{\tem}{\hat{T}}
\newcommand{\lam}{\hat{\Lambda}}
\newcommand{\hint}{\hat{\mathcal{H}}}

\newcommand{\tdf}{\delta \tilde{f}}

\newcommand{\ene}{\hat{\epsilon}}
\newcommand{\intp}{\int_{\hat{p}}} 
\newcommand{\R}{\hat{\mathcal{R}}}
\newcommand{\Ih}{\hat{I}}
\newcommand{\It}{\tilde{I}}
\newcommand{\I}{I}
\newcommand{\calI}{{\hat{\mathcal{I}}}}
\newcommand{\calIeq}{{\mathcal{I}_\mathrm{eq}}}

\newcommand{\J}{\hat{\mathcal{J}}}
\newcommand{\lu}{\hat{l}^\mu}

\newcommand{\xh}{\hat{x}}
\newcommand{\lh}{\hat{l}}
\newcommand{\xu}{\hat{\Xi}^{\mu\nu}}

\newcommand{\she}{\hat{\pi}}
\newcommand{\Pt}{\hat{\mathcal{P}}_\perp}
\newcommand{\Pl}{\hat{\mathcal{P}}_L}

\newcommand{\piti}{{\hat{\tilde{\pi}}}}
\newcommand{\feq}{f_\mathrm{eq}}
\begin{document}
\title{Anisotropic fluid dynamics for Gubser flow}
\date{\today}

\author{M.~Martinez}
\affiliation{Department of Physics, The Ohio State University, Columbus, OH 43210, USA}
\affiliation{Department of Physics, North Carolina State University, Raleigh, NC 27695}

\author{M.~McNelis}
\affiliation{Department of Physics, The Ohio State University, Columbus, OH 43210, USA}

\author{U.~Heinz}
\affiliation{Department of Physics, The Ohio State University, Columbus, OH 43210, USA}

%%%%%%%%%%%%%%%%%%%%%%%%%%%%%%%%%%%%%%%%%%%%%%%%%%%%%%%%%%
\begin{abstract}
Exploring a variety of closing schemes to the infinite hierarchy of momentum moments of the exactly solvable Boltzmann equation for systems undergoing Gubser flow, we study the precision with which the resulting hydrodynamic equations reproduce the exact evolution of hydrodynamic moments of the distribution function. We find that anisotropic hydrodynamics, obtained by expanding the distribution function around a dynamically evolving locally anisotropic background whose evolution is matched to exactly reproduce the macroscopic pressure anisotropy caused by the different longitudinal and transverse expansion rates in Gubser flow, provides the most accurate macroscopic description of the microscopic kinetic evolution. This confirms a similar earlier finding for Bjorken flow [Moln\'ar, Niemi and Rischke, Phys. Rev. D {\bf 94}, 125003 (2016)]. We explain the physics behind this optimal matching procedure and show that one can efficiently correct for a non-optimized matching choice by adding a residual shear stress to the energy-momentum tensor whose evolution is again determined by the Boltzmann equation. Additional insights to guide the optimal choice of a macroscopic anisotropic hydrodynamic framework for strongly-coupled systems that do not admit a microscopic kinetic description are reported.  
\end{abstract}
%%%%%%%%%%%%%%%%%%%%%%%%%%%%%%%%%%%%%%%%%%%%%%%%%%%%%%%%%%


\maketitle

%%%%%%%%%%%%%%%%%%%%%%%%%%%%%%%%%%%%%%%%%%%%%%%%%%%%%%%%%%
\section{Introduction}
\label{sec:intr}
%%%%%%%%%%%%%%%%%%%%%%%%%%%%%%%%%%%%%%%%%%%%%%%%%%%%%%%%%%

Anisotropic hydrodynamics~\cite{Martinez:2010sc,Martinez:2010sd,Florkowski:2010cf,Ryblewski:2010bs,Florkowski:2011jg,Ryblewski:2011aq,Martinez:2012tu,Ryblewski:2012rr,Florkowski:2012as,Florkowski:2013uqa,Bazow:2013ifa,Tinti:2013vba,Florkowski:2014bba,Florkowski:2014txa,Nopoush:2014pfa,Denicol:2014mca,Bazow:2015cha,Nopoush:2015yga,Florkowski:2015cba,Alqahtani:2015qja,Bazow:2015zca,Tinti:2015xra,Bluhm:2015raa,Bluhm:2015bzi,Alqahtani:2016rth,Florkowski:2016kjj,Nopoush:2016qas,Florkowski:2016zsi,Schaefer:2016yzd} is a variant of dissipative (``viscous'') fluid dynamics that addresses the following problem: All real fluids have non-zero mean free paths, resulting in non-zero transport coefficients such as shear and bulk viscosity. In such fluids, hydrodynamic expansion competes with microscopic relaxation processes, keeping an expanding fluid from ever reaching local thermodynamic equilibrium. If the expansion is anisotropic, the different expansion rates along different directions and the associated velocity shear render the momentum distribution of the microscopic constituents anisotropic in the local rest frame (LRF). If this anisotropy becomes large, it manifests itself macroscopically through large shear stresses, resulting in highly anisotropic effective pressures that depend on the expansion rate and its anisotropies.

By Landau matching, the energy and conserved particle densities, $e$ and $\bm{n}$, control the local temperature $T$ and chemical potentials $\bm{\mu}$ of the fluid's constituents in a non-expanding state. The fluid's equation of state $P(e,\bm{n})$ reflects the interactions among the microscopic constituents and is a material property of the fluid. Pressure gradients are the forces driving hydrodynamic expansion, by local acceleration of the fluid. The resulting evolution of the densities of energy, momentum and conserved charges is constrained by conservation laws and causality and therefore happens ``slowly'', i.e. on time scales controlled by spatial pressure and density gradients within the fluid. As long as the microscopic relaxation time scale (i.e.~the mean free time between collisions in a microscopic picture based on colliding particles) is short compared with these macroscopic time scales, the momentum distribution of the microscopic constituents approaches a thermal equilibrium distribution $f_\mathrm{eq}\Bigl(- (u(x)\,{\cdot}\,p)/T(x),\bm{\mu}(x)/T(x)\Bigr)$ in the LRF (which moves with flow four-velocity $u^\mu(x)$ in the global frame), characterized by a local temperature $T(x)$ and chemical potentials $\bm{\mu}(x)$ consistent with the macroscopic energy and conserved particle densities $e(x)$ and $\bm{n}(x)$. Deviations $\delta f(x,p)$ from this isotropic local equilibrium distribution are suppressed by powers of the Knudsen and inverse Reynolds numbers \cite{Denicol:2012cn} which are generically small for a small ratio between the microscopic relaxation and macroscopic hydrodynamic time scales. 

Generically, if expansion ceases, such deviations quickly relax to zero, on a time scale controlled by the microscopic mean free time. For anisotropically expanding systems, however, the local momentum anisotropy persists until its driving force (i.e. the anisotropy of the expansion rate) dies down \cite{Muronga:2004sf, Baier:2006um, Song:2007fn, Song:2007ux}. In spite of the action of microscopic processes that typically erase deviations from local momentum anisotropy on microscopic time scales, the pressure anisotropy thus persists for macroscopic hydrodynamic time scales. Although not constrained by conservation laws (i.e. not a hydrodynamic mode {\it per se}), pressure anisotropies thus evolve slowly, on similar time scales as the energy and particle densities. 

Standard second-order viscous hydrodynamic equations of generalized Israel-Stewart type \cite{Muronga:2001zk,Muronga:2003ta,Heinz:2005bw,Muronga:2006zx,Muronga:2006zw,Baier:2006um,Denicol:2012cn}
are typically derived from momentum moments of the Boltzmann equation, separating ideal and dissipative contributions by expanding the Boltzmann equation around an (isotropic) local equilibrium distribution $f_\mathrm{eq}$. In such an approach, rapidly anisotropically expanding systems lead to large dissipative corrections persisting over long time periods,  challenging the applicability of the expansion on which the approach is based. Anisotropic hydrodynamics expands the Boltzmann equation instead around a deformed momentum distribution in the LRF \cite{Romatschke:2003vc}, with the idea that the deformation parameter $\xi$, characterizing the leading-order term of the distribution function at the microscopic level, should evolve in such a way that, on the macroscopic level, it captures the evolution of the pressure anisotropies caused by the anisotropic expansion. 
   
Over the years many different proposals were made on how to obtain the evolution equation for $\xi$ \cite{Martinez:2010sc,Martinez:2010sd,Florkowski:2010cf,Ryblewski:2010bs,Florkowski:2011jg,Ryblewski:2011aq,Ryblewski:2012rr,Florkowski:2012as,Florkowski:2013uqa,Bazow:2013ifa,Tinti:2013vba,Florkowski:2014bba,Florkowski:2014txa,Nopoush:2014pfa,Denicol:2014mca,Bazow:2015cha,Nopoush:2015yga,Florkowski:2015cba,Alqahtani:2015qja,Bazow:2015zca,Tinti:2015xra,Alqahtani:2016rth,Florkowski:2016kjj,Nopoush:2016qas,Florkowski:2016zsi},
without achieving conceptual clarity. The idea that the deformed leading-order distribution function should be defined such that deviations from the full distribution function (i.e. of the full solution of the Boltzmann equation) are as small as possible \cite{Bazow:2013ifa}, and that residual dissipative effects can therefore be treated perturbatively as in Israel-Stewart theory, is difficult to implement formally. An important insight to circumvent this problem was made by Molnar, Niemi and Rischke~\cite{Molnar:2016vvu,Molnar:2016gwq} who suggested that $\xi$ should be treated as a Lagrange parameter just like the temperature $T$ and chemical potentials $\bm{\mu}$ and chosen such that the leading-order distribution function {\it completely} accounts, at any point in space and time, for the macroscopic longitudinal pressure $P_L$.\footnote{%
	The same suggestion was already made in \cite{Bazow:2013ifa,Bazow:2015zca} whose authors failed, 
	however, to find the correct evolution equations for $\xi$ that would ensure that this matching condition 
	remains preserved over the entire evolution of the fluid.}. 
	A similar matching procedure was previously suggested by 
	Tinti~\cite{Tinti:2015xwa}. Independently Bluhm and Schaefer derived anisotropic non-relativistic hydrodynamics by matching the components of the pressure tensor to the microscopic anisotropy parameter~\cite{Bluhm:2015raa,Bluhm:2015bzi}

This so-called ``$P_L$-matching'' scheme was shown in \cite{Molnar:2016gwq} to yield a form of the anisotropic hydrodynamic equations that almost perfectly reproduces the evolution of all hydrodynamic moments of the distribution function known from the exact solution of the Boltzmann equation in Relaxation Time Approximation (RTA) for a gas of massless Boltzmann particles undergoing Bjorken expansion (i.e. pure boost-invariant longitudinal expansion without transverse gradients) \cite{Baym:1984np,Florkowski:2013lza,
Florkowski:2013lya,Florkowski:2014sfa}. In this highly symmetric scenario the bulk viscous pressure vanishes by conformal symmetry, and the shear stress tensor reduces to a single component whose evolution describes the evolving pressure anisotropy $P_L{-}P_T$, yielding almost exactly the same result as obtained from the corresponding moment of the distribution function whose evolution according to the RTA Boltzmann equation is exactly known. The Boltzmann equation couples the evolution of this shear stress component to higher order (non-hydrodynamic) moments of the distribution function,\footnote{%
	In general the evolution equation of a tensorial moment of the distribution function of rank $r$ 
	couples with higher order moments up to order $r{+}2$. This happens in both standard viscous 
	\cite{Denicol:2012cn} and anisotropic hydrodynamics (see Eqs. (110-112) of Ref.~\cite{Molnar:2016vvu}).}  
and the equation is closed by postulating that the deviation $\delta\tilde f$ of the distribution function from the leading-order term yields a negligible contribution to these moments. The analysis in Ref.~\cite{Molnar:2016gwq} demonstrates that for a system of massless Boltzmann particles undergoing Bjorken flow this approximation is exceedingly accurate, for any value of the microscopic relaxation time (i.e. for any choice of the specific shear viscosity $\eta/s$ (where $\eta$ is the shear viscosity and $s$ the entropy density)) and of the initial momentum-space deformation.

We show in this work that analogous results hold for another situation for which an exact solution of the RTA Boltzmann equation is known \cite{Denicol:2014tha,Denicol:2014xca,Heinz:2015cda}, namely a gas of massless Boltzmann particles undergoing Gubser flow \cite{Gubser:2010ze,Gubser:2010ui} (a flow pattern that combines boost-invariant longitudinal expansion with fast azimuthally symmetric transverse flow). For this case anisotropic hydrodynamics was studied previously by Nopoush, Ryblewski and Strickland (NRS) \cite{Nopoush:2014qba}, however without invoking the $P_L$-matching scheme. We show that using $P_L$ matching improves the accuracy of the anisotropic hydrodynamic evolution, but also that a similarly accurate (although more laborious) scheme can be developed by using the NRS \cite{Nopoush:2014qba} closing prescription and correcting for the resulting slightly inaccurate evolution of the deformation parameter $\xi$ by adding a residual shear stress component that accounts for the non-zero contribution to the shear stress from the deviation $\delta \tilde f$ in this scheme (as was done in \cite{Bazow:2013ifa} for Bjorken flow). The same is not true for standard viscous hydrodynamics (we specifically study its improved DNMR variant \cite{Denicol:2012cn}) which, in spite of accounting for the $\delta f$ deviations from the leading-order (isotropic) distribution, yields rather large deviations from the exact results for the evolution of energy density and shear stress.

Like many studies before, the analysis presented in this work relies strongly on the availability of kinetic theory for the microscopic dynamics. The concept of a momentum deformation parameter $\xi$ makes sense only within kinetic theory. For strongly coupled plasmas, such as the quark-gluon plasma, a microscopic quasiparticle picture may not be available. In such situations, hydrodynamics must be formulated entirely in terms of macroscopic variables, avoiding any recourse to $\xi$ and its dynamics. We use the model at hand, which admits both microscopic and macroscopic descriptions, with analytically known relationships between both, to derive an entirely macroscopic version of anisotropic hydrodynamics with Gubser symmetry. In the microscopic kinetic formulation, the dynamics of the momentum deformation parameter $\xi$ was shown to be controlled not only by the microscopic relaxation time, but also by a higher-order moment of the distribution function which itself also depends on $\xi$, i.e. on the shear stress and thus on the macroscopic dynamic state of the medium. This driving term, which contains both microscopic information about the interactions among the medium constituents (through its dependence on the local pressure) and macroscopic information on the dynamical state of the medium (through its dependence on the local shear stress) is the key new ingredient necessary for an anisotropic hydrodynamic description. While its form is exactly known for the massless gas of Boltzmann particles studied here, it is presently unavailable for the quark-gluon plasma. Until we figure out how to compute it from first principles one will need to model it because without it anisotropic hydrodynamics with a realistic equation of state for hot QCD matter is not well-defined.

Before starting our discussion we introduce our notation. Throughout this work we adopt natural units $\hbar=c=k_B=1$. The metric signature is taken to be ``mostly plus'' $(-,+,+,+)$. In Minkowski space with Milne coordinates $x^{\mu }=(\tau, r, \phi, \eta)$ the line element is given by
\begin{equation}
ds^{2}=g_{\mu \nu }dx^{\mu }dx^{\nu }=-d\tau^{2}+dr^{2}+r^2d\phi^{2}+\tau^2 d\eta^{2}\,,
\end{equation}
where the longitudinal proper time $\tau$, the spacetime rapidity $\eta$, the transverse radius $r$ and the azimuthal angle $\phi$ are given in terms of the Cartesian coordinates ($t,x,y,z$) by
\begin{equation}
\begin{split}
&\tau =\sqrt{t^{2}{-}z^{2}}\,,\hspace{2cm}\eta =\atanh{\left(\frac{z}{t}\right)}\\
&r=\sqrt{x^{2}{+}y^{2}}\,,\hspace{2cm}\phi=\tan^{-1}\left(\frac{y}{x}\right)\,.
\end{split}
\end{equation}
The fluid velocity $u^\mu$ is a time-like normalized vector $u_\mu u^\mu=-1$ which is defined in the Landau Frame, i.e., $u^\mu =T^{\mu\nu}u_\mu/\sqrt{u_\mu T^{\mu\nu}u_\nu}$.

The paper is structured as follows: In Sec.~\ref{sec:Gflow} we briefly review Gubser flow and its associated symmetries. Section~\ref{sec:fluid} is the core of the paper in which we derive two new variants of anisotropic hydrodynamics for Gubser flow and compare their evolution equations with each other, with previously studied different hydrodynamic approximations, and with those arising from the exact solution of the RTA Boltzmann equation. Numerical results from these different approximations are compared and discussed in Sec.~\ref{sec:results}. We summarize our findings in Sec.~\ref{sec:concl}. Some technical details of the derivations of the hydrodynamic evolution equations can be found in the Appendix.

%%%%%%%%%%%%%%%%%%%%%%%%%%%%%%%%%%%%%%%%%%%%%%%%%%%%%%%%
\section{The Gubser flow}
\label{sec:Gflow}
%%%%%%%%%%%%%%%%%%%%%%%%%%%%%%%%%%%%%%%%%%%%%%%%%%%%%%%%

Gubser flow \cite{Gubser:2010ze,Gubser:2010ui} describes conformally symmetric systems that expand azimuthally symmetrically in the transverse plane together with boost-invariant longitudinal expansion. It is most easily described in de Sitter space times a line, $dS_3\otimes R$, where the flow looks static. To proceed from Minkowski space with Milne coordinates to $dS_3\otimes R$ one first performs a Weyl rescaling of the metric,
\begin{equation}
d\hat{s}^2=\frac{ds^2}{\tau^2}
=\frac{-d\tau^2+dr^2+r^2 d\phi^2}{\tau^2}+d\eta^2\,,  
\label{metricdS3R}
\end{equation}
followed by the coordinate transformation $x^\mu=(\tau,r,\phi,\eta)\mapsto \xh^\mu=(\rho,\theta,\phi,\eta)$ where
\begin{subequations}
\label{eq:rhotheta}
\begin{align}
\rho(\tilde\tau,\tilde r)& =-\mathrm{arcsinh}\left( \frac{1-\tilde\tau^2+\tilde r^2}
{2\tilde\tau }\right)\,,
\label{definerho} \\
\theta (\tilde\tau,\tilde r)& =\mathrm{arctan}\left(\frac{2\tilde r}{1+\tilde\tau^2-\tilde r^2}\right) \,,
\label{definetheta}
\end{align}
\end{subequations}
with $\tilde\tau = q\tau$ and $\tilde r = q r$ where $q$ is an arbitrary energy scale that sets the transverse size of the system~\cite{Gubser:2010ze,Gubser:2010ui}. In these coordinates the Weyl transformed line element reads
\begin{equation}
d\hat{s}^{2}=-d\rho ^{2}+\cosh ^{2}\!\rho \left( d\theta ^{2}+\sin
^{2}\theta\, d\phi ^{2}\right) +d\eta ^{2}\,, 
\label{eq:linedS3R}
\end{equation}
with the metric $\g=\mathrm{diag}\bigl(-1,\cosh^2\rho, \cosh^2\rho\sin^2\theta,1\bigr)$ and the square root of the metric determinant $\sqrt{-\hat{g}}=\cosh^2\rho\cos\theta$. The new ``de Sitter time'' coordinate $\rho$ has the range $\rho\in (-\infty,\infty)$ while the new coordinate $ \theta\in (0,2\pi)$ plays the role of an angle. 

The line element (\ref{eq:linedS3R}) is invariant under rotations in the space spanned by $(\theta,\phi)$; the corresponding symmetry group is denoted as $SO(3)_q$~\cite{Gubser:2010ze}. Including the reflection symmetry $\eta\to-\eta$ and longitudinal boost invariance, the line element (\ref{eq:linedS3R}) is invariant under the ``Gubser symmetry'' $\gub$ \cite{Gubser:2010ze}. The only normalized vector that is invariant under this symmetry is $\hat{u}^\mu=(1,0,0,0)$~\cite{Gubser:2010ze,Gubser:2010ui}. This symmetry also implies that macroscopic variables such as the energy density $\ene(\xh)=\ene(\rho)$ depend only on the de Sitter time \cite{Gubser:2010ze, Gubser:2010ui} while phase-space distributions $f(\xh,\pp)=f(\rho,\po,\pp_\eta)$ depend only on $\rho$ \cite{Gubser:2010ze,Gubser:2010ui} and the momentum components $\po=\pp_\theta^2+\pp_\phi^2/\sin^2\theta$ and $\pp_\eta$ conjugate to the coordinates $\theta, \phi$ and $\eta$ \cite{Denicol:2014xca,Denicol:2014tha}.

We denote by variables with a hat all quantities that are expressed in Gubser coordinates $\xh^\mu$. 
%%%%%%%%%%%%%%%%%%%%%%%%%%%%%%%%%%%%%%%%%%%%%%%%%%%%%%%%%%
\section{Fluid dynamics for Gubser flow}
\label{sec:fluid}
%%%%%%%%%%%%%%%%%%%%%%%%%%%%%%%%%%%%%%%%%%%%%%%%%%%%%%%%%%

In this section we review the derivation of fluid dynamical equations for the conformally symmetric Gubser flow from the relativistic Boltzmann equation for a system of massless particles, $\pp^2=0$, using the Relaxation Time Approximation (RTA) for the collision term. In Gubser coordinates, this RTA Boltzmann equation reads \cite{Denicol:2014xca,Denicol:2014tha}
\be
\label{eq:boltzgubeq}
\partial_\rho f(\rho,\po,\pp_\eta)=\frac{1}{\hat{\tau}_r(\rho)}
\biggl(\feq\biggl(\frac{-\hat{u}\,{\cdot}\,\pp}{\tem(\rho)}\biggr) - f(\rho,\po,\pp_\eta)\biggr)
\ee
where $\feq(z){\,=\,}e^{-z}$ is the local thermal equilibrium distribution, 
\be
-\hat{u}\cdot\pp=\sqrt{\hat{p}^2_\Omega / \cosh^2\rho + \hat{p}^2_\eta}
\ee
is the particles' energy in the local rest frame (LRF) of the fluid, and $\tem$ is the temperature. Conformal symmetry requires $\hat{\tau}_r(\rho)=c/\tem(\rho)$ with $c=5\,\bar\eta$ where the specific shear viscosity $\bar\eta\equiv\eta/s$ is the ratio of the shear viscosity and entropy density. The exact solution of Eq.~(\ref{eq:boltzgubeq}) \cite{Denicol:2014xca,Denicol:2014tha} will be given in subsection \ref{subsec:gubsol} below when we need it.

The macroscopic hydrodynamic variables that make up the energy-momentum tensor are obtained as momentum moments of the distribution function,
\be
\label{eq:EMkin}
\tem^{\mu\nu}(\hat{x})=\langle\,\pp^\mu\,\pp^\nu\,\rangle\,,
\ee
where $\langle\mathcal{O}(\hat{x})\rangle\equiv\int_{\pp}\,\mathcal{O}(\hat{x},\hat{p}) f(\hat{x},\hat{p})$, with
\be
\int_{\pp}\equiv\int \frac{d\pp_\theta d\pp_\phi d\pp_\eta}{(2\pi)^3 \sqrt{-\hat{g}}\hat{E}_{\hat{p}}},
\ee
denotes the momentum moment of the phase-space observable $\mathcal{O}(\hat{x},\hat{p})$.\footnote{%
	Note that we define the momentum integration measure in terms of the covariant components of 
	the four-vector $\pp_\mu$ \cite{Denicol:2014xca,Denicol:2014tha,Bazow:2016oky}.
	}
We evaluate these momentum integrals in the LRF where the on-shell energy $\hat{E}_{\hat{p}}=\pp^\rho=-\,\hat{u}\,{\cdot}\,\pp$. Momentum moments of the equilibrium distribution are denoted by $\langle\mathcal{O}(\hat{x},\hat{p})\rangle_{\mathrm{eq}}$.

Using the relation (\ref{eq:EMkin}), any (approximate) dynamical solution for the distribution function can be used to derive a set of hydrodynamic and relaxation equations for the components of the energy-momentum tensor. Taking derivatives of the left hand side of Eq.~(\ref{eq:EMkin}) leads to terms on the right hand side that involve momentum moments of derivatives of the distribution function. These must be evaluated using the Boltzmann equation (\ref{eq:boltzgubeq}) and, in general, couple the hydrodynamic moments in the energy-momentum tensor to higher-order, non-hydrodynamic momentum moments of the distribution function. To close the set of equations one must truncate the momentum hierarchy using some approximation scheme. We now discuss the sets of hydrodynamic equations resulting from several different such closing schemes for systems undergoing Gubser flow. 

%%%%%%%%%%%%%%%%%%%%%%%%%%%%%%%%%%%%%%%%%%%%%%%%%%%%%%%%%%%%%%%%%%%%%%%%
\subsection{Viscous hydrodynamics}
\label{subsec:vischydro}
%%%%%%%%%%%%%%%%%%%%%%%%%%%%%%%%%%%%%%%%%%%%%%%%%%%%%%%%%%%%%%%%%%%%%%%%

For the derivation of standard viscous hydrodynamics from kinetic theory one expands the distribution function around a local equilibrium distribution:
\be
\label{eq:isotropic}
f(\hat{x},\hat{p})=f_\mathrm{eq}\Bigl(\beta_{\hat{u}}(\xh)(-\,\hat{u}(\xh)\,{\cdot}\,\pp)\Bigr) + \delta f (\hat{x},\hat{p})\,,
\ee
where $\beta_{\hat{u}}(\xh)=1/\tem(\xh)$ is the inverse local temperature. The local rest frame energy ${-\,}\hat{u}(\xh){\,{\cdot}\,}\pp$ is isotropic in momentum space. $\delta f$ encodes all deviations from local thermal equilibrium, in particular any local momentum anisotropies caused by anisotropic global expansion. 

In this situation it is convenient to decompose the metric tensor $\hat{g}^{\mu\nu}$ into the locally temporal and spatial projectors, $-\hat{u}^\mu \hat{u}^\nu$ and $\hat{\Delta}^{\mu\nu}=\hat{g}^{\mu\nu}+\hat{u}^\mu \hat{u}^\nu$, respectively, and use these to decompose the particle four-momentum as $\pp^\mu= (-\hat{u}{\cdot}\pp)\,\hat{u}^\mu
+\pp^{\langle\mu\rangle}$. Here $-\hat{u}\cdot\pp$ is the LRF energy, and for any four vector $\hat{A}^\mu$ the transverse projection $\hat{A}^{\langle\mu\rangle}=\hat{\Delta}^{\mu\nu}\hat{A}_\nu$ is a vector that is purely spatial in the LRF. For later use we also introduce $\hat{B}^{\langle\mu\nu\rangle}=\Du_{\alpha\beta}\hat{B}^{\alpha\beta}$ where the double projector $\Du_{\alpha\beta}=(\hat{\Delta}^{\mu}_{\alpha}\hat{\Delta}^{\nu}_{\beta}+\hat{\Delta}^{\mu}_{\beta}\hat{\Delta}^{\nu}_{\alpha}-\frac{2}{3}\Du\hat{\Delta}_{\alpha\beta})/2$ projects the tensor $\hat{B}^{\mu\nu}$ onto its symmetric, traceless and locally purely spatial (i.e. orthogonal to $\hat{u}^\mu(\xh)$) part. 

In the Landau frame the most general form of the energy-momentum tensor is then
\be
\label{eq:enemomvh}
\tem^{\mu\nu}=\ene\,\hat{u}^\mu\hat{u}^\mu+\hat{\mathcal{P}}\hat{\Delta}^{\mu\nu}+\hat{\pi}^{\mu\nu}\,,
\ee
where $\ene$ is the LRF energy density, $\hat{\mathcal{P}}=\hat{P}_0(\ene)+\hat{\Pi}$ is the isotropic pressure, 
and $\she^{\mu\nu}$ is the shear stress tensor. For conformal systems such as the one studied here, the bulk viscous pressure $\hat{\Pi}$ vanishes, and the isotropic pressure $\hat{\mathcal{P}}$ is given by the thermal pressure $\hat{P}_0(\ene)=\ene/3$ obtained from the conformal equation of state. These macroscopic quantities correspond to the following moments of the distribution function $f(\xh,\pp_i)$: 
\begin{subequations}
\begin{align}
&\ene=\hat{u}_\mu\hat{u}_\nu\tem^{\mu\nu}=\langle (\hat{u}\cdot\pp)^2\rangle\,,\\
&\hat{\mathcal{P}}=\frac{1}{3}\hat{\Delta}_{\mu\nu}\tem^{\mu\nu}=\frac{1}{3}\langle\hat{\Delta}_{\mu\nu}\pp^\mu\pp^\nu\rangle\,,\\
\label{eq:shearvisc}
&\she^{\mu\nu}=\tem^{\langle\mu\nu\rangle}=\langle\pp^{\langle\mu}\pp^{\nu\rangle}\rangle\,.
\end{align}
\end{subequations}
In addition to the choice of the LRF velocity as the timelike eigenvector of $\tem^{\mu\nu}$, $\tem^{\mu\nu} \hat{u}_\nu=\ene\,\hat{u}^\mu$, uniqueness of the decomposition (\ref{eq:isotropic}) requires fixing the local inverse temperature $\beta_{\hat{u}}$. This is done through the Landau matching condition \cite{Landaufluid} 
\be
\label{eq:landaumatch}
\ene := \langle (\hat{u}\cdot\pp)^2\rangle_\mathrm{eq} = \ene_\mathrm{eq}(\tem)
         = \frac{3}{\pi^2}\,\tem^4,
\ee 
which ensures that the parameter $\tem$ in $f_\mathrm{eq}$ is adjusted such that $\delta f$ does not contribute to the LRF energy density. As a result, all deviations of the system from local equilibrium are encoded in the shear stress tensor: $\she^{\mu\nu}\equiv\langle\pp^{\langle\mu}\pp^{\nu\rangle}\rangle_{\delta}$ where $\langle\cdots\rangle_{\delta}$ indicates a momentum moment of $\delta f$.

The evolution equation for the energy density $\ene$ is obtained from the time-like (i.e.\ $\hat{u}_\nu$) projection of the energy-momentum conservation law $\hat{D}_\mu \tem^{\mu\nu}=0$ (where $\hat{D}_\mu$ is the covariant derivative). For systems with Gubser symmetry this yields \cite{Gubser:2010ze,Gubser:2010ui}
\be
\label{eq:eqviscener}
\partial _{\rho}	\ene
+ \frac{8}{3}\,\ene\,\tanh\rho
= \hat\pi^{\eta\eta}\tanh\rho\,.
\ee
Such systems have only one independent shear stress component $\hat\pi{\,\equiv\,}\she^{\eta\eta}$. To obtain an evolution equation for $\hat{\pi}$ one can start from the RTA Boltzmann equation and use the method of moments described in Ref.~\cite{Denicol:2012cn} (DNMR).\footnote{%
		For a complete discussion of different methods to derive hydrodynamics from relativistic kinetic theory 
		see Ref.~\cite{Denicol:2014loa}.}
Within the 14-moment approximation one obtains (see Ref.~\cite{Denicol:2014tha}, Appendix A, and Ref.~\cite{Nopoush:2014qba})
\be
\label{eq:eqviscshear}
  \partial _{\rho }\hat\pi+\frac{\hat\pi}{\hat{\tau}_r} + \frac{46}{21}\hat{\pi}\tanh\rho
  = \frac{16}{45}\ene\tanh\rho\,.
\ee
Introducing the normalized shear stress $\hat{\bar\pi}\equiv3\hat\pi/(4\ene)$, the coupled DNMR equations (\ref{eq:eqviscener},\ref{eq:eqviscshear}) can be rewritten as
\begin{subequations}
\label{eq:vhydro}
\begin{align}
\label{eq:vhydro1}
&\partial_\rho\ln\ene = \frac{4}{3}(\hat{\bar\pi}-2)\tanh\rho\,,\\
\label{eq:vhydro2}
&\partial_\rho\hat{\bar\pi} + \frac{\hat{\bar\pi}}{\hat{\tau}_r} =
   \frac{4}{3}\tanh\rho\left( \frac{1}{5} + \frac{5}{14}\hat{\bar\pi} - \hat{\bar\pi}^2 \right).
\end{align}
\end{subequations}

\vspace*{-3mm}
%%%%%%%%%%%%%%%%%%%%%%%%%%%%%%%%%%%%%%%%%%%%%%%%%%%%%%%%%%
\subsection{Anisotropic hydrodynamics}
\label{subsec:ahydro}
%%%%%%%%%%%%%%%%%%%%%%%%%%%%%%%%%%%%%%%%%%%%%%%%%%%%%%%%%%
\vspace*{-2mm}

Anisotropic hydrodynamics generalizes viscous hydrodynamics by allowing for a leading-order dissipative deformation of the distribution function due to anisotropic expansion of the system. If the expansion rate along a certain ``longitudinal'' direction $\hat{l}^\mu$ is much larger or smaller than in the other directions, one can account for this by generalizing the decomposition (\ref{eq:isotropic}), making the leading-order distribution anisotropic in momentum in the LRF by including an additional dependence on the momentum component in the $\hat{l}^\mu$ direction \cite{Bazow:2013ifa,Molnar:2016vvu}:
\be
\label{eq:anisexp}
f(\hat{x},\hat{p})=f_a\left(\beta_{\hat{u}}(-\hat{u}\cdot\pp),\beta_{\hat{l}}(\hat{l}\cdot\pp)\right)+\tdf(\hat{x},\hat{p})\,.
\ee
$\um$, $\hat{l}^\mu$, $\beta_{\hat{u}}$ and $\beta_{\hat{l}}$ are all functions of $\xh$. The parameter $\beta_{\hat{l}}$ parametrizes the strength of the leading-order local momentum anisotropy, and $\delta \tilde{f}$ takes into account residual dissipative corrections. Anisotropic hydrodynamics is expected to be an improvement over viscous hydrodynamics whenever the residual dissipative effects associated with $\delta \tilde{f}$ are smaller than the leading-order dissipative effects manifest in the local momentum anisotropy $\beta_{\hat{l}}$. We demand that in the limit $\beta_{\hat{l}}\to 0$ the anisotropic distribution function $f_a$ reduces to the local equilibrium distribution $f_\mathrm{eq}$ in (\ref{eq:isotropic}).

To account for the effects from the momentum anisotropy of the leading-order distribution function $f_a$ on the structure of the macroscopic energy-momentum tensor it is convenient to perform the tensor decomposition in terms of both the fluid velocity $\um$ and space-like ``longitudinal'' vector $\hat{l}^\mu$ which in the LRF is chosen to point in the $\eta$-direction: $\lu=(0,0,0,1)$. The space orthogonal to these two vectors is spanned by the transverse spatial projector tensor $\xu=\Du-\lu\hat{l}^\nu$ \cite{Huang:2009ue,Huang:2011dc,Gedalin1,Gedalin2,Molnar:2016vvu,Molnar:2016gwq}. The four-momentum can now be decomposed as $\pp^\mu=$ $(-\,\hat{u}\,{\cdot}\,\pp)\,\un + (\hat{l}\,{\cdot}\,\pp)\, \lu+\pp^{\{\mu\}}$ where $\pp^{\{\mu\}}{\,\equiv\,}\xu\pp_\nu$ are the transverse spatial momentum components while $\hat{l}\cdot\pp=\pp_\eta$ is the longitudinal momentum in the LRF.

This leads to the following decomposition of the energy-momentum tensor \eqref{eq:EMkin} in the Landau frame \cite{Molnar:2016vvu}:
\be
\label{eq:MNRemt}
\tmn=\ene\,\um\,\un + \Pl\,\lu\lh^{\nu} + \Pt\,\xu + 2\,\hat{W}^{(\mu}_{\perp\lh}\,\lh^{\nu)} + \piu_\perp,
\ee
with 
\bs
\label{eq:anisodecomp}
\beal
&\ene=\uum\unn\tmn\,\equiv \langle\,(-\,\hat{u}\cdot\pp)^2\,\rangle\,,\\
&\Pl=\lh_\mu\lh_\nu\tmn\,\equiv\langle\,(\hat{l}\cdot\pp)^2\,\rangle\,,\\
&\Pt=\frac{1}{2}\hat{\Xi}_{\mu\nu}\tem^{\mu\nu}\,\equiv\frac{1}{2}\langle\,\hat{\Xi}_{\mu\nu}\pp^\mu\pp^\nu\,\rangle\,,\\
&\hat{W}^{\mu}_{\perp\lh}\,=\hat\Xi^{\mu}_{\alpha}\tem^{\alpha\beta}\lh_\beta\,\equiv\langle\,(\hat{l}\cdot\pp)\,\pp^{\{\mu\}}\rangle\,, \\
&\piu_\perp= \hat\Xi_{\alpha\beta}^{\mu\nu}\tem^{\alpha\beta}\equiv\langle\,\pp^{\{\mu}\pp^{\nu\}}\,\rangle.
\end{align}
\es
In the last line we introduced the notation $B^{\{ \mu\nu\}}\equiv\xu_{\alpha\beta}B^{\alpha\beta}$ where the double projector $\xu_{\alpha\beta}{\,\equiv\,}\frac{1}{2}\big(\hat{\Xi}^\mu_\alpha\hat{\Xi}^\nu_\beta+\hat{\Xi}^\nu_\beta\hat{\Xi}^\mu_\alpha-\xu\hat{\Xi}_{\alpha\beta}\big)$ projects an arbitrary tensor $B^{\mu\nu}$ on its symmetric, traceless and locally spatially transverse part. Comparison of Eqs.~\eqref{eq:shearvisc} and \eqref{eq:MNRemt} shows that the shear stress $\pi^{\mu\nu}$ has been further decomposed as \cite{Molnar:2016vvu,Molnar:2016gwq}
\be
\label{eq:shearrelation}
\hat{\pi}^{\mu\nu}=\she^{\mu\nu}_{\perp}\,+\,2\,\hat{W}^{\left(\mu\right.}_{\perp\lh}\hat{l}^{\left.\nu\right)}\\
+\frac{1}{3} \big(\Pt{-}\Pl\big) \big(\xu{-}2\,\lu\hat{l}^\nu\big).
\ee  

The above decomposition is general. For systems with Gubser symmetry it simplifies considerably. Conformal symmetry requires $\ene=2\Pt{+}\Pl$, corresponding to zero bulk viscous pressure $\Pi$ and a conformal equation of state $\ene=3\hat{P}_0(\ene)=2\Pt{+}\Pl$. The $SO(3)_q\otimes Z(2)$ part of the Gubser symmetry implies
$\she^{\mu\nu}_{\perp}=0=\hat{W}^{\mu}_{\perp\lh}$. This leaves 
\be
\label{eq:anistmn}
\tmn=\ene\,\um\,\un\,+\,\Pl\,\lu\lh^{\nu}\,+\,\Pt\,\xu
\ee
as the most general energy-momentum tensor for systems with Gubser symmetry. The single nonvanishing shear stress component $\hat\pi{\,\equiv\,}\she^{\eta\eta}$ defines the difference between the longitudinal and transverse pressures via  
\be
\label{eq:pihat}
  \Pl-\Pt=\textstyle{\frac{3}{2}}\she
\ee
and the shear stress tensor via
\be
\label{eq:shearreduced}
\begin{split}
\hat{\pi}^{\mu\nu}&=\she\left(\lu\hat{l}^\nu{-}\textstyle{\frac{1}{2}}\xu\right)\,.
\end{split}
\ee  

For the leading-order anisotropic distribution function we use the Romatschke-Strickland (RS) ansatz \cite{Romatschke:2003ms} 
\be
\label{eq:RSansatz}
f_a(\xh,\pp;\hat\Lambda,\xi)=f_\mathrm{eq}\big(E_\mathrm{RS}(\xi)/\hat{\Lambda}\big)
\ee
where $f_\mathrm{eq}(z)=e^{-z}$ is again the Boltzmann distribution but now evaluated for the momentum-anisotropic argument
\be
\label{eq:ERS}
E_\mathrm{RS}(\xi)
  \equiv \sqrt{(\hat{u}\cdot\pp)^2+\xi\,(\hat{l}\cdot\pp)^2} 
  = \sqrt{\frac{\po}{\cosh^2\rho}+(1{+}\xi)\pp_\eta^2}.
\ee
Here $\hat\Lambda$ and $\xi$ are functions of $\xh$. The definition \eqref{eq:ERS} corresponds to parameters $\beta_{\hat{u}}=1/\hat{\Lambda}(\rho)$ and $\beta_{\hat{l}}=\xi\,\beta_{\hat{u}}=\xi(\rho)/\hat{\Lambda}(\rho)$ in \eqref{eq:anisexp}, where we used the fact that Gubser symmetry restricts the space-time dependence of $\hat\Lambda$ and $\xi$ to functions of the de Sitter time $\rho$ only.

Following \cite{Molnar:2016gwq} we define the scalar integrals
\bea
\label{eq:Inlq}
I_{nlq}&\equiv&\bigl\langle(-\hat{u}\cdot\pp)^{n-l-2q} (\hat{l}\cdot\pp)^l (\hat{\Xi}_{\mu\nu}\pp^\mu\pp^\nu)^q\bigr\rangle \nonumber\\
&\equiv& \Ih_{nlq}(\hat{\Lambda},\xi)+\It_{nlq} \,,
\eea
%
%and
%
%\bea
%\label{eq:Inm}
%\calI_{nm}(\hat{\Lambda},\xi)&\equiv&\bigl\langle(-\hat{u}\cdot\pp)^{n} (\hat{l}\cdot\pp)^m \bigr\rangle \nonumber\\
%&\equiv& \calIh_{nm}+\calIt_{nm}
%\eea
%
where the first term on the rhs denotes the leading order contribution from $f_a$ (which depends on the parameters $\hat\Lambda$ and $\xi$) and the second term the subleading contribution from $\delta\tilde{f}$ in Eq.~\eqref{eq:anisexp}. For massless particles the dependences of the leading order $\Ih$ integrals on $\hat\Lambda$ and $\xi$ factorize (see Appendix \ref{app:anisint}, Eq.~\eqref{eq:anisint2}). 

With these definitions the leading order RS distribution function \eqref{eq:RSansatz} contributes to the energy-momentum tensor as follows: 
\be
\label{eq:RStmn}
\tmn_\mathrm{RS}=\ene_\mathrm{RS}\,\um\,\un\,+\,\Pl^\mathrm{RS}\,\lu\lh^{\nu}\,+\,\Pt^\mathrm{RS}\,\xu,
\ee
where 
\bs
\label{eq:RSmacquant}
\begin{align}
\label{eq:RSenergydens}
\ene_\mathrm{RS}&=\left\langle(-\,\hat{u}\cdot\pp)^2\right\rangle_a=\Ih_{200}\big(\hat{\Lambda},\xi\big)\,,\\
\Pl^\mathrm{RS}&=\bigl\langle(\hat{l}\cdot\pp)^2\bigr\rangle_a=\Ih_{220}\big(\hat{\Lambda},\xi\big)\,,\\
\Pt^\mathrm{RS}&=\frac{1}{2}\bigl\langle\hat{\Xi}_{\mu\nu}\pp^\mu\pp^\nu\bigr\rangle_a
%   =\frac{1}{2}\bigl\langle(\hat{u}\cdot\pp)^2{-}(\hat{l}\cdot\pp)^2\bigr\rangle_a\nonumber\\
   =\frac{1}{2}\Ih_{201}\big(\hat{\Lambda},\xi\big).
\end{align}
\es
Using Eq.~\eqref{eq:pihat} and the mass-shell condition in the form $\hat{\Xi}_{\mu\nu}\pp^\mu\pp^\nu=(-\,\hat{u}\cdot\pp)^2-(\hat{l}\cdot\pp)^2$ these relations further imply
\be
\label{eq:RSpi}
  \she_\mathrm{RS}=\bigl\langle(\hat{l}\cdot\pp)^2-\textstyle{\frac{1}{3}}(-\hat{u}\cdot\pp)^2\big\rangle_a
         =\Ih_{220}\big(\hat{\Lambda},\xi\big)-\textstyle{\frac{1}{3}}\Ih_{200}\big(\hat{\Lambda},\xi\big).
\ee
Conformal symmetry implies that the trace of the energy-momentum tensor to vanish exactly, hence
\be
\label{eq:anisotrace}
\ene_\mathrm{RS}= 2\Pt^\mathrm{RS}+\Pl^\mathrm{RS} \,.
\ee

As in the viscous hydrodynamic case, to make the decomposition \eqref{eq:anisexp} of the distribution function unique we need a prescription for the parameters $\beta_{\hat{u}}=1/\hat{\Lambda}$ and $\beta_{\hat{l}}=\xi/\hat{\Lambda}$. For $\hat\Lambda$ we use the Landau matching condition \cite{Martinez:2010sc,Martinez:2010sd}
\be
\label{eq:LMcond}
\ene_\mathrm{RS}(\hat\Lambda,\xi) := \ene_\mathrm{eq}(T)
\quad\Longrightarrow\quad
 \hat{\Lambda} := \frac{\tem}{\big(\R_{200}(\xi)\big)^{1/4}}\,,
\ee
where $\R_{200}(\xi)$ is the $\xi$-dependent part of $\I_{200}(\hat\Lambda,\xi)$ (see Eq.~\eqref{eq:anisint2}). This condition ensures that the first term in the energy-momentum tensor \eqref{eq:anistmn} receives no contribution from the residual deviation $\delta\tilde{f}$ in Eq.~\eqref{eq:anisexp}, $\ene_\mathrm{RS}=\ene$. 

For the second parameter $\xi$ a number of different prescriptions have been proposed \cite{Martinez:2010sc, Martinez:2010sd, Florkowski:2010cf, Ryblewski:2010bs, Florkowski:2011jg, Ryblewski:2011aq, Ryblewski:2012rr, Florkowski:2012as, Florkowski:2013uqa, Bazow:2013ifa, Tinti:2013vba, Florkowski:2014bba,Florkowski:2014txa,Nopoush:2014pfa,Denicol:2014mca,Bazow:2015cha,Nopoush:2015yga,Florkowski:2015cba,Alqahtani:2015qja,Bazow:2015zca,Tinti:2015xra,Alqahtani:2016rth,Florkowski:2016kjj,Nopoush:2016qas,Florkowski:2016zsi}. They correspond to different ways of splitting the distribution function $f$ into a leading order term $f_a$ and a residual deviation $\delta\tilde{f}$. Since (as we will see) the terms in the energy momentum tensor arising from $f_a$ and from $\delta\tilde{f}$ are treated differently in the derivation of equations of motion, these different prescriptions lead to different anisotropic hydrodynamic equations. Different anisotropic hydrodynamic approximations can, in this sense, be characterized by the different ways the anisotropy parameter $\xi(\rho)$ evolves. In the following subsections we discuss three different such possibilities. It is worth noting that Eq.~\eqref{eq:LMcond} implies that for a given value of the energy density $\ene$ (which fixes $\tem$) the value of $\hat\Lambda$ depends on $\xi$ and is thus sensitive to the different evolutions of $\xi$ in these different versions of anisotropic hydrodynamics.  

\vspace*{-3mm}
%%%%%%%%%%%%%%%%%%%%%%%%%%%%%%%%%%%%%%%%%%%%%%%%%%%%%%%%%%
\subsection{$\mathcal{P}_L$ matching}
\label{subsec:plmatching}
%%%%%%%%%%%%%%%%%%%%%%%%%%%%%%%%%%%%%%%%%%%%%%%%%%%%%%%%%%
\vspace*{-2mm}

We start with the simplest and, as it turns out, most effective prescription for $\xi$, the $\Pl$-matching scheme first proposed in \cite{Bazow:2013ifa} and recently successfully implemented for systems undergoing Bjorken flow 
by Molnar {\it et\ al.}\ \cite{Molnar:2016vvu,Molnar:2016gwq}.\footnote{%
	We note that the $\Pl$-matching condition \eqref{eq:PLmatch} below is a special case of a more general 
	prescription proposed earlier by Tinti \cite{Tinti:2015xwa} for (3+1)-dimensional anisotropic hydrodynamics 
	which uses a generalized Romatschke-Strickland form for the leading order distribution function $f_a$
	that is flexible enough to capture {\em all} components of the energy-momentum tensor (by appropriately 
	matching its parameters), i.e. the residual deviation $\delta\tilde{f}$ of the distribution function contributes
	{\em nothing} to $T^{\mu\nu}$. By adapting the derivations in \cite{Tinti:2015xwa} to Bjorken- and 
	Gubser-symmetric situations we checked that they lead to the same results as those reported in 
	Ref.~\cite{Molnar:2016gwq} and in this subsection, respectively, for the $\Pl$-matching scheme.}
This prescription considers $\xi$ as a parameter that represents on the microscopic level the macroscopic longitudinal pressure $\Pl$ (or, more precisely, the shear stress component $\she$ that is responsible for the longitudinal-transverse pressure difference $\Pl{-}\Pt$), in very much the same way as the temperature $\tem$ represents the energy density $\ene$. It should therefore be fixed by a corresponding ``Landau matching condition'' that adjusts the value of $\xi$ in $f_a$ such that $f_a$ fully captures all contributions to $\Pl$ (or, equivalently, to $\she$), i.e. $\Pl$ and $\she$ receive no contribution from the residual deviation $\delta\tilde{f}$ of the distribution function:
\be
\label{eq:PLmatch}
\hat{\mathcal{P}}_L\equiv\hat{\mathcal{P}}^\mathrm{RS}_L(\Lambda,\xi) 
\quad \Longleftrightarrow\quad \she=\she_\mathrm{RS},\ \hat{\tilde{\pi}}=0 .
\ee
With this additional matching condition the leading order energy-momentum tensor \eqref{eq:RStmn} and the full energy-momentum tensor \eqref{eq:anistmn} become identical, i.e. the residual deviation $\delta\tilde{f}$ does not contribute at all to $\tem^{\mu\nu}$, but only to higher-order non-hydrodynamic momentum moments of the distribution function.

Using the identities $\Pl=\hat{P}_0(\ene) +\she=\ene/3+\she$ and $\Pt=\hat{P}_0(\ene) -\she/2 =\ene/3-\she/2$ in \eqref{eq:anistmn} the energy conservation law takes exactly the same form as in viscous hydrodynamics, Eqs.~\eqref{eq:eqviscener} or \eqref{eq:vhydro1}.

The equation of motion for $\she$ is most easily obtained by using the RTA Boltzmann equation to derive an equation of motion for $\Pl$ and then use $\she=\Pl-\hat{P}_0=\Pl-\ene/3$. A straightforward calculation yields
\bea
\label{eq:MNRlongpress}
 \parho\hat{\mathcal{P}}_L 
 &=&\parho\left(\int_\pp \pvv^2\, f \right)
\\\nonumber
 &=&-\frac{\Pl-\frac{1}{3}\ene}{\hat{\tau}_{r}}\,-\left(\Pl+\I_{240}\right)\tanh\rho\,,
\eea
where we used the mass-shell condition $\pp\cdot\pp=-(\ph)^2+\pp_\eta^2 + \po/\cosh^2\rho=0$ to eliminate $\po$ from the integration measure as well as the Landau matching conditions for $\ene$ and $\Pl$. Equation \eqref{eq:MNRlongpress} is not closed because $\I_{240}$ still involves an integral over the full distribution function; to close the equation we can approximate it by dropping the $\delta\tilde{f}$ contribution to $\I_{240}$ \cite{Molnar:2016gwq} by replacing $\I_{240} \to \Ih_{240}(\hat\Lambda,\xi)$. Substituting this approximation together with $\Pl=\ene/3+\she$ into Eq.~\eqref{eq:MNRlongpress} yields the following equation for $\she$:
\be
\label{eq:PLshear}
  \partial_\rho\she+\frac{\she}{\hat{\tau}_r} + \tanh\rho\Bigl(\frac{4}{3}\she + \Ih_{240}(\hat\Lambda,\xi)\Bigr)
  = \frac{5}{9}\ene\tanh\rho\,.
\ee
This should be compared with Eq.~\eqref{eq:eqviscshear} in viscous hydrodynamics. As before this equation can be rewritten for the normalized shear stress $\hat{\bar{\pi}}=3\she/(4\ene)$,
\be
\label{eq:ahydroPl}
   \partial_\rho\hat{\bar\pi} + \frac{\hat{\bar\pi}}{\hat{\tau}_r} =
   \frac{4}{3}\tanh\rho\left(\frac{5}{16} + \hat{\bar\pi} - \hat{\bar\pi}^2 - \frac{9}{16}\mathcal{F}(\hat{\bar\pi})\right) ,
\ee
which should be compared with Eq.~\eqref{eq:vhydro2}. Here
\be
\label{eq:F}
   \mathcal{F}(\hat{\bar\pi}) \equiv \frac{\R_{240}\bigl(\xi(\hat{\bar\pi})\bigr)}{\R_{200}\bigl(\xi(\hat{\bar\pi})\bigr)},
\ee
where $\xi(\hat{\bar\pi})$ is the inverse of the function
\bea
\label{eq:pibarxi}
   \hat{\bar\pi}(\xi) &=& \frac{3\she}{4\ene}=\frac{3\,\Ih_{220}{-}\Ih_{200}}{4\,\Ih_{200}}
%\nonumber\\
   = \frac{1}{4}\left(\frac{3\,\R_{220}(\xi)}{\R_{200}(\xi)}-1\right)\qquad
\eea
(see Eqs.~\eqref{eq:RSmacquant} and \eqref{eq:RSpi}), with the $\R$ functions given in \eqref{eq:Rfunctions}. Equation \eqref{eq:pibarxi} can be used to compute (by numerical inversion) the de Sitter time evolution of $\xi$ from the solution $\hat{\bar\pi}(\rho)$ of the anisotropic hydrodynamic equations.

Note that the coupled anisotropic hydrodynamic equations \eqref{eq:vhydro1} and \eqref{eq:ahydroPl} are formulated entirely in terms of the macroscopic hydrodynamic variables $\ene$ and $\hat{\bar\pi}$, without taking recourse to the microscopic parameters $\hat\Lambda$ and $\xi$. This was achieved by exploiting the factorization \eqref{eq:anisint2} of the $\hat\Lambda$ and $\xi$ dependencies in the modified thermal integrals over the RS distribution function $\I_{nlq}$ that holds for massless particles. The only differences between the DNMR equations \eqref{eq:vhydro} and anisotropic hydrodynamics are somewhat different factors multiplying the constant and linear terms in $\hat{\bar\pi}$ and the appearance of the function $\mathcal{F}(\hat{\bar\pi})$ on the rhs of Eq.~\eqref{eq:ahydroPl}. For nonconformal theories the function $\mathcal{F}$ (which arises from the term $\Ih_{240}$ in \eqref{eq:PLshear}) depends on both $\ene$ and $\hat{\bar\pi}$.  $\mathcal{F}(\ene,\hat{\bar\pi})$ serves as an additional driving force for the shear stress that arises from the competition between the microscopic interactions driving the fluid towards local momentum isotropy and the anisotropic expansion driving the system away from it. As such, it depends on both the intrinsic properties of the medium (reflected in its dependence on $\ene$ or transverse temperature $\hat\Lambda$) and its actual dynamical state (reflected in its dependence on $\hat{\bar\pi}$ or the momentum anisotropy parameter $\xi$). While $\mathcal{F}(\ene,\hat{\bar\pi})$ is analytically known for the non-interacting massless Boltzmann gas with Gubser symmetry studied here, it is not obvious how to calculate it in QCD from first principles for a system undergoing arbitrary anisotropic expansion. We regard $\mathcal{F}(\ene,\hat{\bar\pi})$ as an additional, anisotropic driving force which for interacting theories must be suitably parametrized until a way of computing it from first principles has been found.
 
For comparison with the following subsections we also present the evolution equations in terms of the microscopic parameters $\hat{\Lambda}$ and $\xi$. Using the energy and $\Pl$ matching conditions together with Eqs.~(\ref{eq:RSmacquant},\ref{eq:RSpi}), we rewrite the energy conservation law \eqref{eq:eqviscener} in terms of the scalar integrals \eqref{eq:Inlq}:
\be
\label{eq:lambdaPL}
  \partial_\rho\Ih_{200}(\hat{\Lambda},\xi) +
  \tanh\rho\,\left(3\,\Ih_{200}(\hat{\Lambda},\xi) - \Ih_{220}(\hat{\Lambda},\xi)\right)=0.
\ee
Chain rule differentiation $\partial_\rho\Ih_{200}(\hat{\Lambda},\xi)=(\partial_{\hat\Lambda}\Ih_{200})\,\partial_\rho\hat\Lambda$ $+\, (\partial_{\xi}\Ih_{200})\,\partial_\rho\xi$ turns this into an equation that couples the $\rho$ derivatives of $\hat\Lambda$ and $\xi$. They can be uncoupled by using Eq.~(\ref{eq:ahydroPl}), rewritten (with $\hat{\bar{\pi}}(\xi)$ from \eqref{eq:pibarxi}) as
\be
\label{eq:drhoxi}
  \partial_\rho\xi + \frac{1}{\hat{\tau}_r}\,\frac{\hat{\bar{\pi}}(\xi)} {\partial_\xi\hat{\bar{\pi}}(\xi)} = - 2\tanh\rho\, (1{+}\xi).
\ee
The rhs of this equation (which, due to subtle cancellations, turns out to be surprisingly simple when compared with that of Eq.~(\ref{eq:ahydroPl})!) controls the free-streaming ($\hat{\tau}_r{\,\to\,}\infty$) evolution of the anisotropy parameter $\xi$, and thereby the late-time behavior of the hydrodynamic quantities.\footnote{%
	The factor $2\tanh\rho$ on the rhs represents the scalar expansion rate  $\hat\theta=\hat{D}{\,\cdot\,}\hat{u}$ 
	of the Gubser flow \cite{Gubser:2010ze,Gubser:2010ui,Denicol:2014tha}.}

%%%%%%%%%%%%%%%%%%%%%%%%%%%%%%%%%%%%%%%%%%%%%%%%%%%%%%%%%%
\subsection{The NRS prescription}
\label{subsec:LONRS}
%%%%%%%%%%%%%%%%%%%%%%%%%%%%%%%%%%%%%%%%%%%%%%%%%%%%%%%%%%

Anisotropic hydrodynamics for Gubser symmetric systems was discussed previously by Nopoush, Ryblewski and Strickland (NRS)~\cite{Nopoush:2014qba}. Instead of using $\Pl$ matching, they considered a linear combination of third order moments $\mathcal{I}^{\mu\nu\lambda}\equiv \int_\pp \pp^\mu \pp^\nu \pp^\lambda f$ of the distribution function, specifically the combination 
\be
\label{eq:NRS1}
  \mathcal{I} \equiv (1{+}\xi)\I_{320} - \textstyle{\frac{1}{2}} \I_{301} \equiv \calI(\hat\Lambda,\xi) +\tilde{\mathcal{I}},
\ee
where on the right hand side we split $\mathcal{I}$ into its leading order contribution $\calI(\hat\Lambda,\xi)$ (from $f_a$ in Eqs.~(\ref{eq:RSansatz},\ref{eq:ERS})) and the residual $\tilde{\mathcal{I}}$ (from $\delta\tilde{f}$), in analogy to Eq.~(\ref{eq:Inlq}). Using the RTA Boltzmann equation, NRS derived the following equation of motion for $\calI(\hat\Lambda,\xi)$:
\be
\label{eq:NRSsecondmom}
   \partial_\rho \calI \,-\, \Ih_{320}\,\partial_\rho \xi \,+\, \frac{\calI{-}\calIeq}{\hat{\tau}_r} 
   = - 2\tanh\rho \left(\calI{-}\frac{1}{2} \Ih_{301}\right) ,
\ee
where $\calIeq$ is the corresponding combination of third order moments of the local equilibrium distribution $\feq$. In this derivation (which is accurate to leading order in the expansion of the distribution function around $f_a$) all contributions to $\mathcal{I}$ from $\delta\tilde{f}$ are neglected.

By dimensional analysis and thanks to the factorization of the $\hat\Lambda$ and $\xi$ dependencies for massless particles, $\calI$ is proportional to $\ene^{5/4}$. Normalizing Eq.~\eqref{eq:NRSsecondmom} by $\ene^{5/4}$ one obtains after some algebra \cite{Nopoush:2014qba} the following equation of motion for the anisotropy parameter $\xi$:
\be
\label{eq:NRSxidot}
  \partial_\rho\xi  + \frac{\xi(1+\xi)^{3/2}\,\R^{5/4}_{200}(\xi)}{\hat{\tau}_r}
  = - 2\tanh\rho\, (1{+}\xi).
\ee
Eqs.~(\ref{eq:lambdaPL},\ref{eq:NRSxidot}) constitute the NRS scheme. This should be compared with Eq.~(\ref{eq:lambdaPL},\ref{eq:drhoxi}) in the $\Pl$-matching scheme. Clearly the anisotropy parameter $\xi$ evolves differently in the NRS and $\Pl$-matching schemes.

%%%%%%%%%%%%%%%%%%%%%%%%%%%%%%%%%%%%%%%%%%%%%%%%%%%%%%%%%%
\subsection{Residual dissipative corrections to the NRS prescription}
\label{subsec:NLONRS}
%%%%%%%%%%%%%%%%%%%%%%%%%%%%%%%%%%%%%%%%%%%%%%%%%%%%%%%%%%

Since with the NRS prescription $\xi$ evolves differently than in the $\Pl$-matching scheme, the moment $\int_\pp \pp^2_\eta\, f_a(\hat\Lambda,\xi)$ no longer fully matches the macroscopic longitudinal pressure $\Pl$. Instead, the latter receives an additional contribution $\piti$ from $\delta\tilde{f}$ which has to make up for the missing piece:
\be
\label{eq:PLNRS} 
  \Pl=\Pl^\mathrm{RS} + \piti=\Ih_{220}(\hat\Lambda,\xi)+\int_\pp \pp_\eta^2\, \delta\tilde{f}.
\ee
This residual shear stress correction on the longitudinal pressure was not taken into account in \cite{Nopoush:2014qba}, i.e. NRS continued to solve for the energy conservation law the equation \eqref{eq:lambdaPL} which only accounts for the leading order contribution to $\Pl$ from an incorrectly matched anisotropic distribution function $f_a$. Inclusion of the residual shear stress from $\delta\tilde{f}$ modifies the energy conservation law as follows:
\be
\label{eq:NLONRSener2}
\partial_\rho\Ih_{200}(\hat{\Lambda},\xi)
+ \tanh\rho\left(3\,\Ih_{200}(\hat{\Lambda},\xi){-}\Ih_{220}(\hat{\Lambda},\xi)\right)
= \piti\tanh\rho.
\ee
Obviously, now an additional equation of motion for $\piti$ is needed.\footnote{%
	This closely mirrors the situation in Ref.~\cite{Bazow:2013ifa} where (for a system with Bjorken
	symmetry) the $\xi$ evolution was obtained from the zeroth moment of the RTA Boltzmann 
	equation, which then necessitated the inclusion and propagation of a residual shear stress
	component $\hat{\tilde\pi}$ to correctly evolve the longitudinal pressure. For the case of Gubser 
	symmetry we found that using the zeroth moment of the RTA Boltzmann equation as an
	equation of motion for $\xi$ causes the evolution of the total shear stress $\hat\pi$ to fail, by 
	developing a singularity whenever $\xi$ approaches a zero crossing. We have not been able to find 
	a remedy for this unphysical behavior associated with the use of the zeroth moment for Gubser flow. 
	We view this as an illustration of the principle that the zeroth moment of the Boltzmann equation
	should be used to determine the evolution of the (non-equilibrium) chemical potential of the 
	leading order distribution function (which in the present work was set to zero) rather than that of
	the momentum anisotropy parameter.}  
It is derived from the RTA Boltzmann equation using the standard procedure (see e.g.~\cite{Denicol:2012cn}): 
Writing $f{\,=\,}f_a{+}\delta\tilde{f}$, Eq.~\eqref{eq:boltzgubeq} gives the following evolution equation for $\delta\tilde{f}$:
\be
\label{eq:deltafeq}
   \partial_\rho \tdf = - \frac{f_a+\tdf-\feq}{\hat{\tau}_{r}} - \partial_\rho f_a
\ee
where from Eqs.~(\ref{eq:RSansatz},\ref{eq:ERS})
\be
\label{eq:der-frs}
\besp
   \partial_\rho f_a = \biggl(\frac{\partial_\rho{\hat{\Lambda}}}{\hat{\Lambda}^2} E_\mathrm{RS} 
   + \frac{1}{\hat{\Lambda}\,E_\mathrm{RS}} \Big(\!\tanh\rho \frac{\hat{p}^2_\Omega}
   {\cosh^2\rho} - \frac{\partial_\rho{\xi}}{2}\,\hat{p}^2_\eta \Big)\! \biggr) f_a.
\end{split}
\ee
With this the residual shear stress $\piti\equiv\piti^{\eta\eta}$ evolves as
\begin{widetext}
\be
\label{eq:ressheareq}
\besp
  \parho\piti =\parho\,\left(\intp\,\pp^{\langle\eta}\pp^{\eta\rangle}\,\tdf\,\right)
  =&-\frac{\hat{\pi}_\mathrm{RS}{+}\piti}{\hat{\tau}_{r}}-\tanh\rho\left(\frac{4}{3}\piti{+}\It_{240}\right)
    -\frac{\partial_\rho{\hat{\Lambda}}}{\hat{\Lambda}^2}\,\left(\hint_{221}{-}\frac{1}{3}\hint_{201}\right)	
\\
    &-\frac{\tanh\rho}{\hat{\Lambda}}\left(\frac{4}{3}\hint_{42-1}{-}\hint_{44-1}-\frac{1}{3}\hint_{40-1}\right)
    +\frac{\partial_\rho{\xi} }{2\hat{\Lambda}}\,\left(\hint_{44-1}{-}\frac{1}{3}\hint_{42-1}\right).
\end{split}
\ee
The anisotropic integrals $\hint_{nlr}$ are defined in Appendix~\ref{app:anisint}, Eq.~\eqref{eq:Hfunc}, and $\It_{240}$ is the $\tdf$ contribution to $I_{240}$.
\end{widetext}
Equation~\eqref{eq:ressheareq} is exact but not closed without an approximation for the residual deviation $\tdf$ in the integral $\It_{240}$.\footnote{%
	We found that simply setting $\It_{240}$ to zero on the rhs of Eq.~\eqref{eq:ressheareq} is 
	inconsistent with keeping $\piti$ nonzero and leads to an incorrect asymptotic behavior of 
	the total shear stress as large $\rho$.} 
We here use the 14-moment approximation for $\tdf$. Its specific form for our present situation is derived in Appendix~\ref{app:14mom}:
%\vspace*{-4mm}
%
\bea
\label{eq:14gradgub}
\tdf_{14} &=&\left[ \hat{\alpha} + \hat{\beta}\,\hat{u}\cdot\pp
                  + \frac{4}{3}\hat{\omega}(\hat{u}\cdot\pp)^2 \right.
\nonumber\\
        &&\ \left.
        + \frac{1}{2} \hat{\omega}_{\langle\eta\eta\rangle} \Big(3(\hat{l}\cdot\pp)^2{-}(\hat{u}\cdot\pp)^2 \Big) \right] 
         f_a\,,
\eea
where (with $\alpha_{\tilde{\pi}}(\xi)$, $\beta_{\tilde{\pi}}(\xi)$, $\gamma_{\tilde{\pi}}(\xi)$ and $\kappa_{\tilde{\pi}}(\xi)$ given in Eqs.~\eqref{eq:alphapi}-\eqref{eq:kappapi})\\
\bs
\label{eq:14gradcoefgub}
\begin{align}
  \hat{\alpha} &= \frac{\hat{\tilde{\pi}} \, \alpha_{\tilde{\pi}}(\xi)}{\J_{2}(\hat{\Lambda})}\,,
  \qquad
  \hat{\beta} = \frac{\hat{\tilde{\pi}} \, \beta_{\tilde{\pi}}(\xi)}{\J_{3}(\hat{\Lambda})} \,,
\\
  \hat{\omega} &= \frac{\hat{\tilde{\pi}} \, \gamma_{\tilde{\pi}}(\xi)}{\J_{4}(\hat{\Lambda})} \,,
  \qquad
  \hat{\omega}_{\langle\eta\eta\rangle} 
  = \frac{\hat{\tilde{\pi}} \, \kappa_{\tilde{\pi}}(\xi)}{\J_{4}(\hat{\Lambda})}.
\end{align}
\es

Using this approximation for $\tdf$ the calculation of the term $\It_{240}$ in Eq.~\eqref{eq:ressheareq} is straightforward. Some algebra leads then to the following closed evolution equation for the residual shear stress $\piti$: 
\begin{widetext}
\be
\label{eq:ressheareq2}
\besp
\parho\piti =
    & - \frac{\hat{\pi}_\mathrm{RS}{+}\piti}{\hat{\tau}_{r}}
         - \tanh\rho\left[\frac{4}{3}\piti + \hat{\alpha}\,\Ih_{240} - \hat{\beta}\,\Ih_{340} 
                              + \frac{4}{3}\hat{\omega}\,\Ih_{440} 
                              + \frac{1}{2}\hat{\omega}_{\langle\eta\eta\rangle}\left(3\,\Ih_{460}{-}\Ih_{440}\right)\,\right]
\\
   & - \frac{\partial_\rho{\hat{\Lambda}}}{\hat{\Lambda}^2}\,\left(\hint_{221}{-}\frac{1}{3}\hint_{201}\right)	
      - \frac{\tanh\rho}{\hat{\Lambda}}\left(\frac{4}{3}\hint_{42-1}{-}\hint_{44-1}-\frac{1}{3}\hint_{40-1}\right)
      + \frac{\partial_\rho{\xi} }{2\hat{\Lambda}}\,\left(\hint_{44-1}{-}\frac{1}{3}\hint_{42-1}\right).
\end{split}
\ee
\end{widetext}
The evolution equations \eqref{eq:NLONRSener2}, \eqref{eq:NRSxidot} and \eqref{eq:ressheareq2} define the NLO NRS prescription for anisotropic hydrodynamics for Gubser flow. Note that the NLO residual dissipative corrections do not affect the evolution of the momentum anisotropy parameter $\xi$ which remains the same as in the leading-order NRS treatment of Ref.~\cite{Nopoush:2014qba}. The NLO corrections only modify the evolution of the energy density $\ene$ and pressure anisotropy $\Pl-\Pt\sim\hat{\pi}=\hat{\pi}_\mathrm{RS}{+}\piti$.

We conclude this subsection by noting that, due to the appearance of non-hydrodynamic higher order moments $\Ih_{nlq}$ in the NRS and NLO NRS prescriptions, the latter lead to evolution equations that explicitly refer to the evolution of the microscopic momentum anisotropy parameter $\xi$ and cannot be formulated purely macroscopically. This sets them apart from the viscous hydrodynamic and $\Pl$-matching anisotropic hydrodynamic formulations. It is therefore not clear how to generalize the NRS and NLO NRS prescriptions to strongly coupled situations where a microscopic kinetic description in terms of quasiparticle distribution functions is not possible.\hfil\break

%%%%%%%%%%%%%%%%%%%%%%%%%%%%%%%%%%%%%%%%%%%%%%%%%%%%%%%%%%%%%%%%%%%%%%%% 
\subsection{Exact solution of the Boltzmann equation}
\label{subsec:gubsol}
%%%%%%%%%%%%%%%%%%%%%%%%%%%%%%%%%%%%%%%%%%%%%%%%%%%%%%%%%%%%%%%%%%%%%%%% 

For Gubser flow, the RTA Boltzmann equation \eqref{eq:boltzgubeq} is solved exactly by \cite{Denicol:2014xca,Denicol:2014tha} 
\be
\label{eq:exsol}
\begin{split}
f_{ex}&(\rho;\po,\pvv) =D(\rho,\rho_0) f_0(\rho_0;\po,\pvv)\,\\
&+\,\frac{1}{c}\int_{\rho_0}^\rho d\rho'\,D(\rho,\rho')\,\hat{T}(\rho')\, f_{eq}\big(\hat{E}_\pp(\rho')/\tem(\rho')\big)\,,
\end{split}
\ee
with damping function $D(\rho,\rho_0){\,=\,}\exp\!\left[-\frac{1}{c}\int_{\rho_0}^\rho d\rho' \,\hat{T}(\rho')\right]$ and initial condition $f_0$ at de Sitter time $\rho_0$ for which we take an RS distribution \eqref{eq:RSansatz} with initial transverse temperature $\hat{\Lambda}_0$ and initial momentum anisotropy $\xi_0$:
\be
\label{eq:RS}
\!\!\!\!\!
f_{0}(\rho_0;\po,\pvv)= 
\exp\left(-\frac{1}{\hat{\Lambda}_0}\sqrt{\frac{\po}{{\cosh^2}\rho_0}+(1{+}\xi_0)\pp_\eta^2}\,\right)\!.
\ee
For $\xi_0\to0$, $f_0\to\feq$. $\feq\big(\hat{E}_\pp(\rho')/\tem(\rho')\big)$ in \eqref{eq:exsol} is the thermal equilibrium distribution at time $\rho'$.

%%%%%%%%%%%%%%%%%%%%%%%%%% Fig 1 %%%%%%%%%%%%%%%%%%%%%%%
\begin{figure*}[t]
\centering
\includegraphics[width=\linewidth]{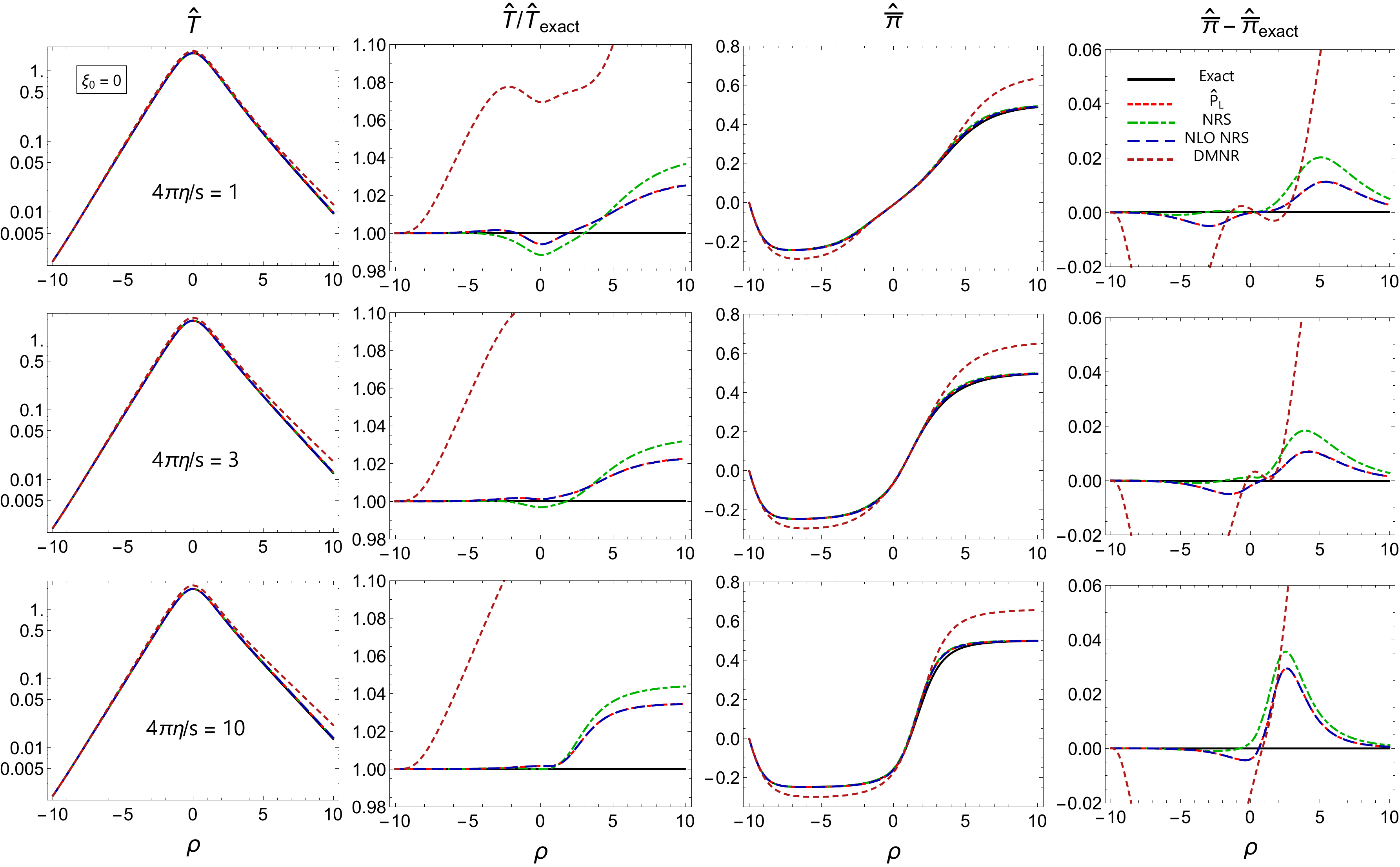}
\caption{(Color online)
de Sitter time evolution of the temperature $\tem$ and the normalized shear stress $\hat{\bar{\pi}}$ for the exact solution of the RTA Boltzmann equation (black solid lines) and four different hydrodynamic approximations: second-order viscous hydrodynamics (DNMR theory, short-dashed magenta lines), anisotropic hydrodynamics with $\Pl$-matching (dotted red lines), leading order anisotropic hydrodynamics in the NRS scheme (dash-dotted green lines), and next-to-leading order anisotropic hydrodynamics in the NRS scheme amended by residual viscous corrections (long-dashed blue lines). For the initial momentum distribution we here assumed isotropy, i.e. $\xi_0=0$. The top, middle and bottom rows of panels correspond to specific shear viscosity $4\pi\eta/s=1$, 3, and 10, respectively. The four columns of plots show, from left to right, the $\rho$ evolution of the temperature $\tem$, of the ratio between its hydrodynamic and exact kinetic evolution $\tem/\tem_\mathrm{exact}$, of the normalized shear stress $\hat{\bar\pi}$, and of the difference between its hydrodynamic and exact kinetic evolution $\hat{\bar\pi}{-}\hat{\bar\pi}_\mathrm{exact}$.
\label{F1}
}
\end{figure*}
%%%%%%%%%%%%%%%%%%%%%%%%%%%%%%%%%%%%%%%%%%%%%%%%%%%%

From the exact solution for $f$ one can find the exact evolution of the energy density $\ene{\,=\,}\Ih_{200}$ and shear stress $\hat{\pi}{\,=\,}\Ih_{220}{-}\frac{1}{3}\Ih_{200}$ \cite{Denicol:2014xca,Denicol:2014tha}: 
\begin{widetext}
\be
\label{eq:exactene}
\ene(\rho) 
        =D(\rho,\rho_0)\Big(\frac{\cosh\rho_0}{\cosh\rho}\Big)^{\!4}\,
        \ene_\mathrm{RS}\big(\lam_0, \xi_\mathrm{FS}(\rho;\rho_0,\xi_0)\big)
        + \frac{1}{c}\int_{\rho_0}^\rho d\rho'\,D(\rho,\rho')\,\hat{T}(\rho')\Big(\frac{\cosh\rho^\prime}{\cosh\rho}
           \Big)^{\!4}\,
         \ene_\mathrm{RS}\big(\tem(\rho^\prime), \xi_\mathrm{FS}(\rho;\rho^\prime,0)\big), 
\ee
\be
\label{eq:exactpi}
\pit(\rho)
        =D(\rho,\rho_0)\Big(\frac{\cosh\rho_0}{\cosh\rho}\Big)^{\!4}\,
        \pit_\mathrm{RS} \big(\lam_0, \xi_\mathrm{FS}(\rho;\rho_0,\xi_0)\big)
        + \frac{1}{c}\int_{\rho_0}^\rho d\rho'\,D(\rho,\rho')\,\hat{T}(\rho')\Big(\frac{\cosh\rho^\prime}{\cosh\rho}
            \Big)^{\!4}\,\pit_\mathrm{RS}\big(\tem(\rho^\prime), \xi_\mathrm{FS}(\rho;\rho^\prime,0)\big).
\ee
\end{widetext}
Here we defined $\xi_\mathrm{FS}(\rho;\rho_\alpha,\xi_\alpha) = -1 + (1+\xi_\alpha)\big(\frac{\cosh\rho_\alpha}{\cosh\rho}\big)^2$. The temperature for the equilibrium distribution is obtained from the energy density by Landau matching, $\ene_\mathrm{RS}(\lam,\xi)=(3/\pi^2)\,\lam^4\R_{200}(\xi)=3\,\tem^4/\pi^2$. When discussing the results in the following section we usually plot the evolution of the temperature rather than the energy density. The specific shear viscosity $\bar\eta=\eta/s$ of the system is tuned by varying $c=5\bar\eta$. We solve the above integral equations for $\ene$ (or $\tem$) and $\pit$ numerically using the method described in Refs.~\cite{Banerjee:1989by,Florkowski:2013lza,Florkowski:2013lya,Denicol:2014tha}.

%%%%%%%%%%%%%%%%%%%%%%%%%% Fig 2 %%%%%%%%%%%%%%%%%%%%%%%
\begin{figure*}[ht]
\centering
\includegraphics[width=\linewidth]{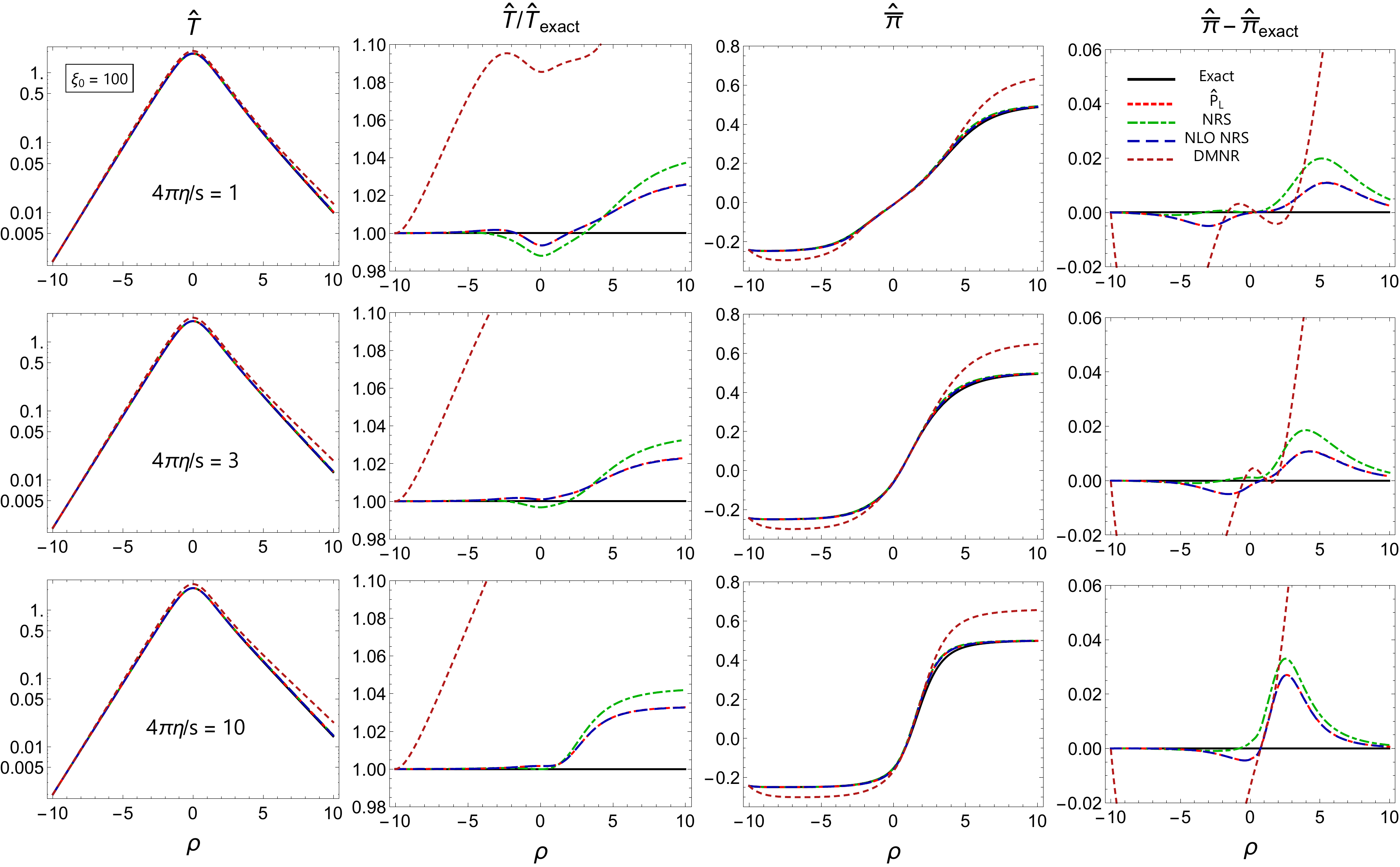}
\caption{(Color online) Same as Fig.~\ref{F1}, but for an initially highly oblate momentum distribution, $\xi_0=100$. 
\label{F2}
}
\end{figure*}
%%%%%%%%%%%%%%%%%%%%%%%%%%%%%%%%%%%%%%%%%%%%%%%%%%%%

%%%%%%%%%%%%%%%%%%%%%%%%%% Fig 3 %%%%%%%%%%%%%%%%%%%%%%%
\begin{figure*}[ht]
\centering
\includegraphics[width=\linewidth]{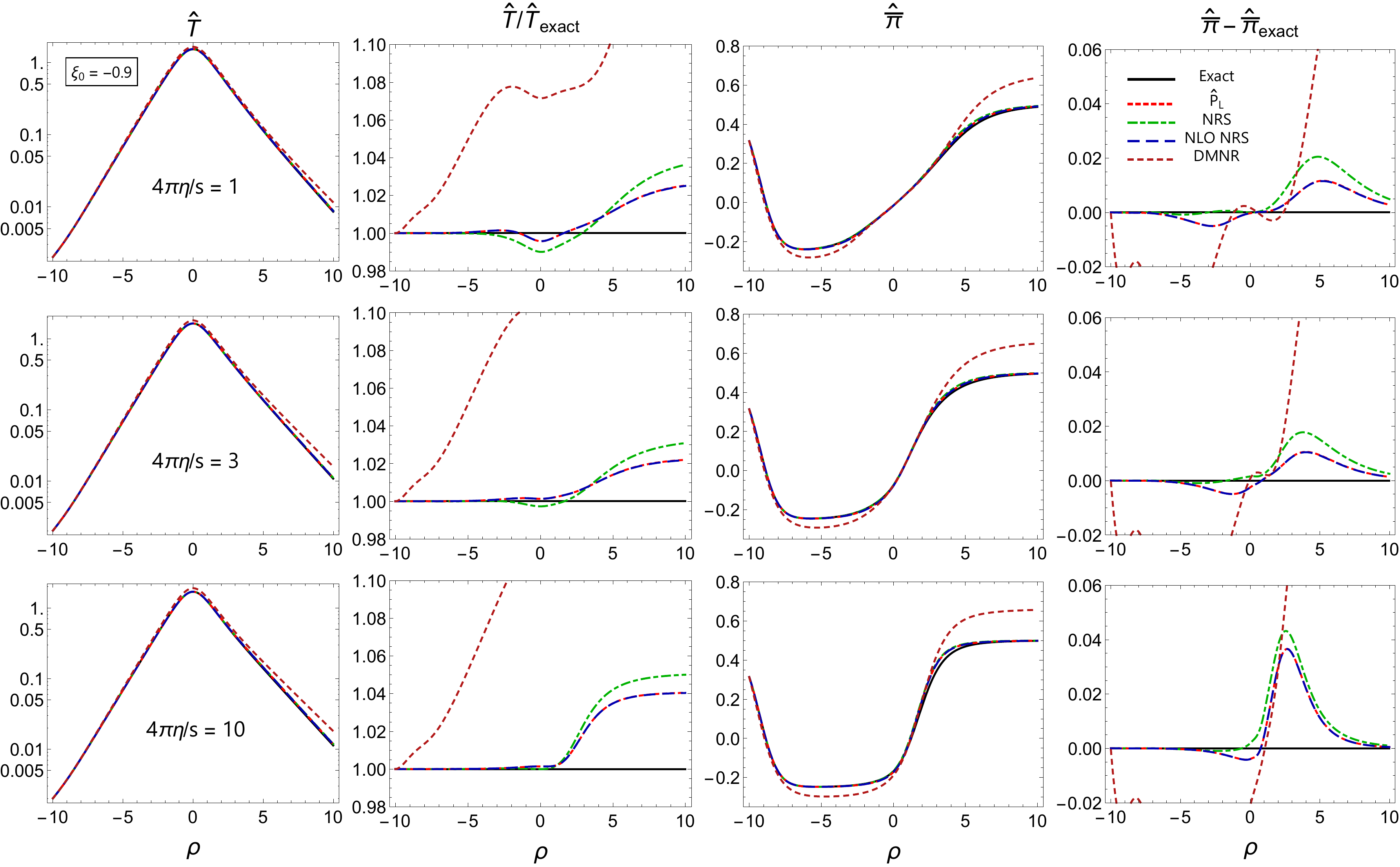}
\caption{(Color online) Same as Fig.~\ref{F1}, but for an initially highly prolate momentum distribution, $\xi_0=-0.9$. 
\label{F3}
}
\end{figure*}
%%%%%%%%%%%%%%%%%%%%%%%%%%%%%%%%%%%%%%%%%%%%%%%%%%%%

%%%%%%%%%%%%%%%%%%%%%%%%%%%%%%%%%%%%%%%%%%%%%%%%%%%%%%%%%%%%%%%%%%%%%%%%
\section{Results}
\label{sec:results}
%%%%%%%%%%%%%%%%%%%%%%%%%%%%%%%%%%%%%%%%%%%%%%%%%%%%%%%%%%%%%%%%%%%%%%%%

In this section we compare the numerical results obtained from the five different approaches discussed in the preceding section (viscous hydrodynamics (DNMR), anisotropic hydrodynamics based on the $\Pl$-matching, NRS and NLO NRS prescriptions, and the exact solution of the Boltzmann equation) for the evolution of the temperature, normalized shear stress and momentum anisotropy parameter. For the $\Pl$-matching scheme and exact solution of the Boltzmann equation we solve equations for the macroscopic shear stress; we convert the shear stress to a momentum anisotropy parameter using the matching condition \eqref{eq:pibarxi}.

For the benefit of the reader we summarize the set of equations that are being solved in each of the five cases:
\begin{itemize}
\item
{\sl Viscous hydrodynamics (DNMR):} Equations~(\ref{eq:vhydro}) for $\ene$ and $\hat{\bar\pi}$, together with $\ene=3\,\tem^4/\pi^2$. In this approach $\xi\equiv0$.
\item
{\sl Anisotropic hydrodynamics with $\Pl$-matching:} Equations \eqref{eq:vhydro1}, \eqref{eq:ahydroPl} together with $\ene{\,=\,}3\,\tem^4/\pi^2$ and Eq.~\eqref{eq:pibarxi} to obtain $\tem$ and $\xi$ from $\ene$ and $\hat{\bar\pi}$.
\item
{\sl Anisotropic hydrodynamics, NRS prescription:} Equations \eqref{eq:lambdaPL} and \eqref{eq:NRSxidot} for $\lam$ and $\xi$, from which we get $\tem$ and $\hat{\bar\pi}$ using Eqs.~\eqref{eq:LMcond} and \eqref{eq:pibarxi}.
\item
{\sl Anisotropic hydrodynamics, NLO NRS prescription:} Equations \eqref{eq:NRSxidot} and \eqref{eq:NLONRSener2} for $\lam$ and $\xi$, together with Eq.~\eqref{eq:ressheareq2} for the residual shear stress $\piti$ and Eqs.~\eqref{eq:LMcond} and \eqref{eq:pibarxi} to get $\tem$ and the leading order $\hat{\bar\pi}(\xi)$ from $\lam$ and $\xi$. 
\item
{\sl Exact solution of the RTA Boltzmann equation:} Equations \eqref{eq:exactene} (rewritten in terms of $\tem$) and \eqref{eq:exactpi}, together with Eq.~\eqref{eq:pibarxi} to get $\xi$ from $\hat{\bar\pi}$.
\end{itemize}

All five models are started at an initial de Sitter time $\rho_0{\,=\,}-10$ with initial temperature $\tem_0 = 0.002$ and run for three choices of the specific shear viscosity, $4\pi\eta/s=1,\,3,$ and 10 (top, middle and bottom row of panels in Figs.~\ref{F1}-\ref{F3}), and three choices of the initial momentum anisotropy, $\xi_0=0,\,100,$ and -0.9 (shown in Figs.~\ref{F1}, \ref{F2}, and \ref{F3}, respectively). We now proceed to discuss these results in detail.

%%%%%%%%%%%%%%%%%%%%%%%%%%%%%%%%%%%%%%%%%%%%%%%%%%%%%%%%%%
\subsection{Evolution of temperature and shear stress}
\label{sec:tempi}
%%%%%%%%%%%%%%%%%%%%%%%%%%%%%%%%%%%%%%%%%%%%%%%%%%%%%%%%%%

Figures~\ref{F1} - \ref{F3} show the de Sitter time evolution of the temperature $\tem$ and of the normalized shear stress $\hat{\bar\pi}$, in the first and third column in absolute values, and in the second and fourth column relative to the exact solution of the RTA Boltzmann equation. It is obvious that all three anisotropic hydrodynamic schemes studied here vastly outperform standard second-order viscous hydrodynamics. Over the range of de Sitter times studied here, anisotropic hydrodynamics with $\Pl$-matching never deviates from the exact solution by more than a few percent. The results from the leading-order NRS scheme (for which the momentum anisotropy parameter is not matched to the macroscopic pressure anisotropy) performs slightly worse, but not dramatically so. Once the residual shear stress caused by the non-optimal $\xi$-evolution in this approach is added to the formalism at next-to-leading order, the evolution of both $\tem$ and $\hat{\bar\pi}$ agrees almost perfectly with that in the $\Pl$-matching scheme.  

For all of the anisotropic hydrodynamic schemes the normalized shear stress $\hat{\bar\pi}$ correctly approaches the asymptotic free-streaming value $\frac{1}{2}$ predicted by the RTA Boltzmann equation \cite{Denicol:2014tha}, in contrast to standard second-order viscous fluid dynamics; the asymptotic temperature lies a few percent above the exact value. The asymptotic behavior at large de Sitter times is almost independent of the initial momentum anisotropy $\xi_0$: due to the rapid expansion the system quickly loses its memory of the initial state. As the specific shear viscosity increases (corresponding to increasing values of the microscopic relaxation time), the deviations between the hydrodynamic evolution of the macroscopic observables and that extracted from the exact solution of the Boltzmann equation grow a bit. However, even for $\eta/s$ values 10 times larger than the ``minimal'' value $1/(4\pi)$ \cite{Policastro:2001yc,Kovtun:2004de} the deviations from the exact solution stay below 4\% for both the temperature ratio $\tem/\tem_{ex}$ and the difference $\hat{\bar\pi} - \hat{\bar\pi}_{ex}$ as long as the $\Pl$-matching scheme is employed.

%%%%%%%%%%%%%%%%%%%%%%%%%%%%%%%%%%%%%%%%%%%%%%%%%%%%%%%%%%
\subsection{Evolution of the momentum anisotropy $\xi$}
\label{sec:xi}
%%%%%%%%%%%%%%%%%%%%%%%%%%%%%%%%%%%%%%%%%%%%%%%%%%%%%%%%%%

%%%%%%%%%%%%%%%%%%%%%%%%%% Fig 4 %%%%%%%%%%%%%%%%%%%%%%%
\begin{figure*}[ht]
\centering
\includegraphics[width=0.8\linewidth]{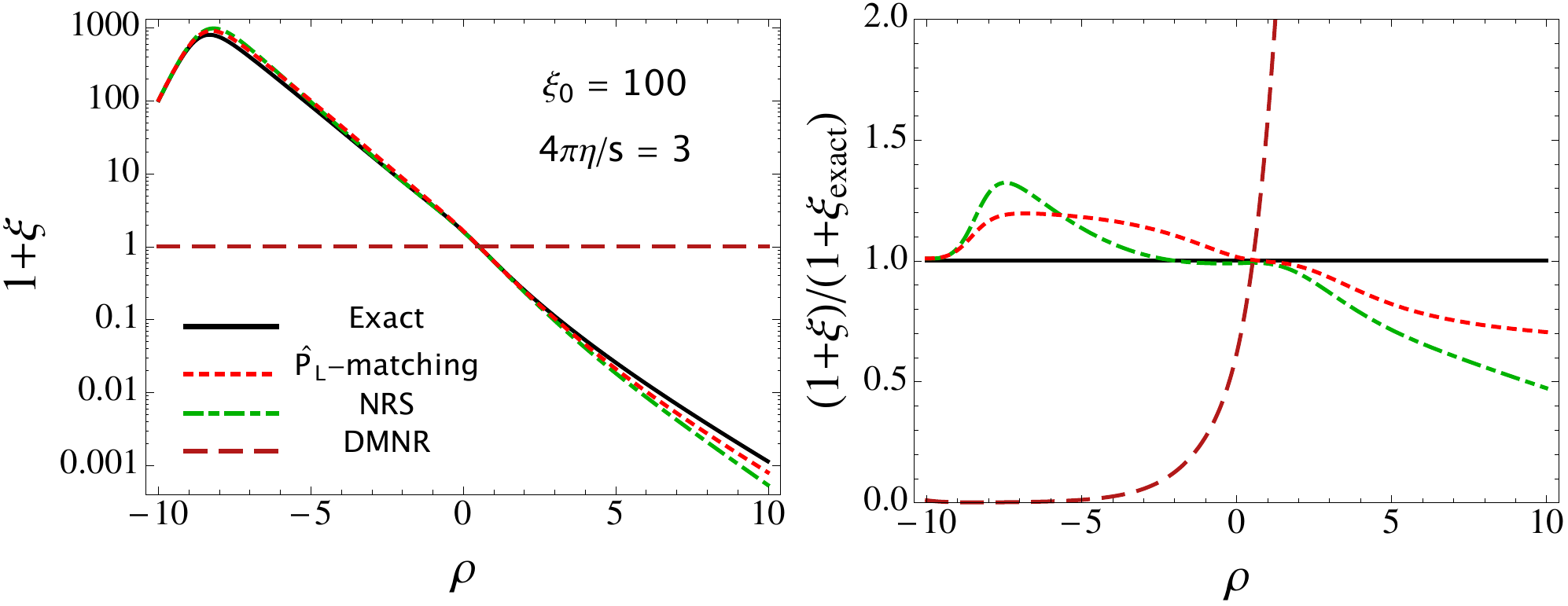}
\caption{(Color online) Comparison of the de Sitter time evolution of the momentum anisotropy parameter $\xi$ corresponding to the exact solution of the RTA Boltzmann equation and the various hydrodynamic approximations discussed in this work, for an initial value $\xi_0=100$ and a specific shear viscosity $4\pi\eta/s=3$. The different approaches are plotted using the same line styles as in Figs.~\ref{F1} - \ref{F3}. See text for discussion.
\label{F4}
}
\end{figure*}
%%%%%%%%%%%%%%%%%%%%%%%%%%%%%%%%%%%%%%%%%%%%%%%%%%%%

As described at the beginning of this section, for the system discussed in this paper which has a microscopic kinetic description in terms of a distribution function of massless particles, the normalized macroscopic pressure anisotropy $\hat{\bar\pi}=(\Pl{-}\Pt)/(2\ene)$ can be related to the microscopic momentum anisotropy parameter $\xi$.\footnote{%
	To be precise, for the NLO NRS scheme, $\xi$ is obtained from the leading order contribution 
	to the pressure anisotropy.
	}
In Fig.~\ref{F4} we compare the de Sitter time evolution of this parameter (offset by 1 because $1\,{+}\,\xi$ is always positive) in the left panel. Whereas second-order viscous hydrodynamics (DNMR theory) is based on the assumption $\xi{\,=\,}0$ at all times, the anisotropic hydrodynamic schemes all show very large and strongly time-dependent deviations from this value, in agreement with the prediction from the exact solution of the Boltzmann equation. In Fig.~\ref{F4} we show the case of an initially strongly oblate momentum distribution $\xi_0{\,=\,}100$, but we have also studied other initial conditions with smaller initial $\xi_0$ value and find that in all cases $1\,{+}\,\xi$ initially grows exponentially with $\rho$ until it hits the falling part of the curve shown in the left panel of Fig.~\ref{F4} and then approximately follows that curve. At large $\rho$ all curves approach $\xi\to-1$, due to an exponential growth of the Knudsen number Kn$=\hat{\theta}\,\hat{\tau}_r$ where $\hat{\theta}$ is the scalar expansion rate. For Gubser flow, Kn$\,\sim\sinh^{2/3}\rho$ at large de Sitter times \cite{Denicol:2014tha}, caused by the exponential decrease of the temperature. This leads to the eventual freeze-out of the momentum distribution due to lack of collisions, driving the system towards an asymptotic free-streaming state \cite{Denicol:2014tha}.

In detail, the $\xi$-evolution differs slightly among the two different closing schemes for anisotropic hydrodynamics studied in this work and between them and the exact solution from the Boltzmann equation. To emphasize these differences we plot in the right panel of Fig.~\ref{F4} the ratio $(1\,{+}\,\xi)/(1\,{+}\,\xi_\mathrm{exact})$. While the anisotropic hydrodynamic schemes are clearly a huge improvement over standard second-order viscous hydrodynamics (DNMR), their $\xi$ evolutions still differ from the exact solution by several 10\% when judged by this ratio. However, for the $\Pl$-matching scheme the evolution of $\xi$ clearly follows the exact trajectory more closely than for the NRS scheme. 
  
The observant reader may notice that all of the curves appear to cross zero near $\rho=0$ (i.e. close to the de Sitter time at which dynamics switches from contraction to expansion and the scalar expansion rate passes through zero). This naive impression turns out to be misleading: Careful inspection shows that the crossing occurs at $\rho>0$ and moves to larger $\rho$ values as the shear viscosity (or microscopic relaxation time) increases. It is also slightly different for the exact solution of the Boltzmann equation and the two different anisotropic hydrodynamic approximation schemes. 

\vspace*{-2mm}
%%%%%%%%%%%%%%%%%%%%%%%%%%%%%%%%%%%%%%%%%%%%%%%%%%%%%%%%%%%%%%%%%%%%%%%%
\section{Conclusions}
\label{sec:concl}
%%%%%%%%%%%%%%%%%%%%%%%%%%%%%%%%%%%%%%%%%%%%%%%%%%%%%%%%%%%%%%%%%%%%%%%%
\vspace*{-3mm}

In this work we studied conformal systems undergoing Gubser flow that admit simultaneously a kinetic description via the RTA Boltzmann equation and a macroscopic hydrodynamic description. We reviewed the standard second-order viscous hydrodynamic formulation and compared it with three different variants of anisotropic hydrodynamics, implementing different closing schemes. In the comparison with the dynamical evolution of the hydrodynamic moments of the distribution function obtained from the exact solution of the Boltzmann equation for this situation, anisotropic hydrodynamics with the $\Pl$-matching scheme revealed itself as the most accurate macroscopic approximation, similar to what was observed in \cite{Molnar:2016gwq} for Bjorken flow. In this approach the evolution of the microscopic momentum anisotropy parameter $\xi$ is matched to the expansion-driven macroscopic pressure anisotropy $\Pl{-}\Pt$. We also showed that an alternate closing scheme proposed in \cite{Nopoush:2014qba} where the $\xi$ evolution does not match the pressure anisotropy can be made equally accurate, albeit with more work by solving a larger set of equations, by accounting for the residual shear stress associated with the non-optimal $\xi$ evolution (NLO NRS scheme). Both procedures lead to descriptions that deviate from the exact results by at most a few percent.

We emphasize that, in contrast to this NLO NRS scheme, anisotropic hydrodynamics with $\Pl$-matching can be formulated entirely macroscopically, without explicitly referring to the underlying kinetic description and its parameters. In particular, the momentum anisotropy parameter $\xi$ can be equivalently replaced by the normalized shear stress $\hat{\bar\pi}$. This puts the anisotropic hydrodynamic treatment on the same footing as second-order viscous hydrodynamics: it can be generalized from conformal systems with Gubser symmetry to non-conformal systems undergoing arbitrary expansion by simply swapping out the ideal conformal equation of state $\ene=3\hat{\mathcal{P}}$ for a realistic EOS (such as the lattice QCD EOS for an expanding quark-gluon plasma). The only non-trivial step in this generalization is the choice of the function $\mathcal{F}(\hat{\bar\pi})$ in Eq.~\eqref{eq:ahydroPl} which describes a nonlinear coupling that describes how the evolution of the shear stress $\hat{\bar\pi}$ in a given fluid cell depends not only on the temperature but also on the pressure anisotropy already established by the degree of anisotropy of the expansion rate in that cell. In the general nonconformal case $\mathcal{F}$ will additionally depend on the energy density, $\mathcal{F}(\ene,\hat{\bar\pi})$. While it can be computed analytically for the Gubser symmetric situation studied here, we do not know yet how to determine this function from first principles for other types of liquids. More research is needed to identify suitable parametrizations for $\mathcal{F}(\ene,\hat{\bar\pi})$ that can be used in heavy-ion collisions. 

It has been stated in \cite{Romatschke:2016hle} that hydrodynamics is valid as long as non-hydrodynamic modes can be safely ignored, but breaks down when this is no longer the case. The evolution of the pressure anisotropy $P_L{-}P_T$ is non-hydrodynamic, in the sense that it manifestly depends on the microscopic relaxation time and is not directly controlled by a conservation law. It clearly plays a crucial and non-negligible role in relativistic heavy-ion collisions, without invalidating the hydrodynamic approach. It requires microscopic knowledge of QCD (just like knowledge of the equation of state (EOS) is required for ideal fluid dynamics), but unlike the EOS it also depends on the macroscopic dynamical state of the fluid. 

%%%%%%%%%%%%%%%%%%%%%%%%%%%%%%%%%%%%%%%%%%%%%%%%%%%%%%%%
\acknowledgments
%%%%%%%%%%%%%%%%%%%%%%%%%%%%%%%%%%%%%%%%%%%%%%%%%%%%%%%%

We thank M. Nopoush, R. Ryblewski and M. Strickland for clarifying discussions of their work reported in Ref.~\cite{Nopoush:2014qba}. We also thank D.~Bazow and J. Noronha for useful comments on the project during its early stages. This work was supported by the U.S. Department of Energy, Office of Science, Office for Nuclear Physics under Awards \rm{DE-SC0004286} and \rm{DE-FG02-03ER41260}. UH and MM also acknowledge generous support through a bilateral scientific exchange program between the Office of Sponsored Research at The Ohio State University and FAPESP which facilitated a visit by MM with J. Noronha's group at the University of Sao Paulo (USP) that led to fruitful discussions of this work.  

\appendix
%%%%%%%%%%%%%%%%%%%%%%%%%%%%%%%%%%%%%%%%%%%%%%%%%%%%%%%%
\section{Anisotropic integrals}
\label{app:anisint}
%%%%%%%%%%%%%%%%%%%%%%%%%%%%%%%%%%%%%%%%%%%%%%%%%%%%%%%%

Here we calculate the anisotropic thermal integrals $\Ih_{nlq}$ and $\hint_{nlr}$ that appear in this paper. First we define\newpage
\bea
\label{eq:anisint}
   \Ih_{nlq}\big(\hat{\Lambda},\xi\big)
   &=&\langle\,(-\hat{u}\cdot\pp)^{n-l-2q}\,(\hat{l}\cdot\pp)^l\,(\hat{\Xi}_{\mu\nu}\pp^\mu\pp^\nu)^q\,\rangle_a
   \nonumber\\
   &=& \int_\pp \, (\pp^\rho)^{n-l-2q}\,\pp_\eta^l\, \left(\frac{\po}{\cosh^2\rho}\right)^q\,f_a.
\eea
The change of variables
\bs
\label{eq:changecoor}
\beal
&\frac{\pth}{\cosh\rho}=\lambda\,\sin\alpha\,\cos\beta\,,\\
&\frac{\pph}{\cosh\rho\,\sin\theta}=\lambda\,\sin\alpha\,\sin\beta\,,\\
&\pp_\eta=\lambda\,(1+\xi)^{-1/2}\cos\alpha
\end{align}
\es
leads to the following factorization of the integral~\eqref{eq:anisint}:
\bea
\label{eq:anisint2}
\Ih_{nlq}\big(\hat{\Lambda},\xi \big)
   &=&\,\J_{n}\big(\hat{\Lambda}\big)\,\R_{nlq}\left(\xi\right)\,,
\\
\label{eq:Jfunc}
\J_{n}\big(\hat{\Lambda}\big)
  &=& \int_0^\infty \frac{d\lambda}{2\pi^2}\,\lambda^{n+1}\, e^{-\lambda/\hat{\Lambda}} 
    = \frac{(n{+}1)!}{2\pi^2}\,\hat{\Lambda}^{n+2},
\\
\label{eq:Rfunc}
\nonumber
\R_{nlq}\left(\xi\right)
  &=&\frac{1}{2(1{+}\xi)^{\left(n-2q\right)/2}} \int_{-1}^1 d\cos\alpha\,\left(\sin\alpha\right)^{2q}\,\left(\cos\alpha\right)^l\,
\\
  &&\times\left[(1+\xi)\,\sin^2\alpha\,+\,\cos^2\alpha\right]^{(n-l-2q-1)/2}\!\!\!\!.\ \ 
\eea
The $\R_{nlq}$ functions needed in this paper are
\begin{widetext}
\bs
\label{eq:Rfunctions}
\beal
\label{eq:R}
\R_{200}(\xi)&=\frac{1}{2}\left(
\frac{1}{1+\xi}+\frac{{\arctan}\sqrt{\xi}}
{\sqrt{\xi}}
\right) &
\partial_\xi \,\R_{200}(\xi)&=\frac{1}{4\xi}\left(
\frac{1-\xi}{(1+\xi)^2}-\frac{{\arctan}\sqrt{\xi}}
{\sqrt{\xi}}
\right) \\
\R_{201}(\xi)&=\frac{1}{2\xi}\left( 1+ (\xi{-}1) \frac{{\arctan}\sqrt{\xi}}{\sqrt{\xi}} \right) &
\R_{220}(\xi)&=\frac{1}{2\xi}\left(
-\frac{1}{1+\xi}+\frac{{\arctan}\sqrt{\xi}}
{\sqrt{\xi}} 
\right) \\
\R_{240}(\xi)&=\frac{1}{2\xi^2}\left(
\frac{3+2\xi}{1+\xi}-3\frac{{\arctan}\sqrt{\xi}}
{\sqrt{\xi}} 
\right) &
\R_{301}(\xi)&=\frac{2}{3(1+\xi)^{3/2}} \\
\R_{320}(\xi)&=\frac{1}{3(1+\xi)^{1/2}} &
 \R_{340}(\xi)&=\frac{1}{\xi^2(1+\xi)^{1/2}}\left(
-\frac{(3+4\xi)}{3(1+\xi)}+\frac{\text{arctanh}\sqrt{\frac{\xi}{1+\xi}}}
{\sqrt{\frac{\xi}{1+\xi}}}
\right) \\
\R_{400}(\xi)&=\frac{1}{8}\left(
\frac{5+3\xi}{(1+\xi)^2}+3\frac{{\arctan}\sqrt{\xi}}
{\sqrt{\xi}} 
\right) &
\R_{420}(\xi)&=\frac{1}{8\xi}\left(
\frac{\xi - 1}{(1+\xi)^2}+\frac{{\arctan}\sqrt{\xi}}
{\sqrt{\xi}}
\right)\\
\R_{440}(\xi)&=\frac{1}{8\xi^2}\left(
-\frac{3+5\xi}{(1+\xi)^2}+3\frac{{\arctan}\sqrt{\xi}}
{\sqrt{\xi}} 
\right)  &
\R_{460}(\xi)&=\frac{1}{8\xi^3}\left(
\frac{8\xi^2+25\xi+15}{(1+\xi)^2}-15\frac{{\arctan}\sqrt{\xi}}
{\sqrt{\xi}} 
\right) 
\end{align}
\es
\end{widetext}%
In Sect.~\ref{sec:fluid} we needed the moments $\Ih_{nlq}^\mathrm{eq}$ associated with the equilibrium distribution function. They are obtained as the isotropic limit of Eq.~\eqref{eq:anisint2}:
\be
\label{eq:equilmom}
\Ih_{nlq}^\mathrm{eq}(\tem)\equiv \lim_{\xi\to 0} \Ih_{nlq}\big(\hat{\Lambda},\xi \big)= \Ih_{nlq}\big(\tem,0)\,.
\ee
Finally, we define the anisotropic integrals\newpage
\be
\label{eq:Hfunc}
  \hint_{nlr}\big(\hat{\Lambda},\xi\big) 
  = \big\langle(-\hat{u}\cdot\pp)^{n-l}(\hat{l}\cdot\pp)^l\,E_\mathrm{RS}^r \big\rangle_a 
  = \int_\pp\! \hat{E}_\pp^{n{-}l}\,\pp_\eta^l\,E_\mathrm{RS}^r\,f_a\,.
\ee
With the change of variables~\eqref{eq:changecoor} one can show that
\be
\label{eq:Hfunc2}
\hint_{nlr}\big(\hat{\Lambda},\xi\big) = \J_{n+r}\big(\hat{\Lambda}\big)\,\R_{nl0}\left(\xi\right).
\ee

\begin{widetext}
%
%%%%%%%%%%%%%%%%%%%%%%%%%%%%%%%%%%%%%%%%%%%%%%%%%%%%%%%%%%%%%%%%%%%%%%%%%%%%%%
\section{14-moment approximation in anisotropic viscous hydrodynamics}
\label{app:14mom}
%%%%%%%%%%%%%%%%%%%%%%%%%%%%%%%%%%%%%%%%%%%%%%%%%%%%%%%%%%%%%%%%%%%%%%%%%%%%%%

In this section we present the calculation of the four coefficients which enter into the 14-moment approximation \eqref{eq:14gradgub} for $\tdf$. We begin with the most general 14-moment ansatz for anisotropic fluids
\cite{Bazow:2013ifa} and decompose it in LRF coordinates for Gubser flow, using the notation $\hat{E}_\pp \equiv \pp^\rho=-\hat{u}\cdot\pp$:
\be
\label{eq:14grad}
%\besp
\tdf  \approx \tdf_{14} 
= \left[\hat{a} + \hat{\beta}_\mu \hat{p}^\mu + \hat{\omega}_{\mu\nu} \hat{p}^\mu \hat{p}^\nu \right] f_a
= \left[ \hat{\alpha} - \hat{\beta} \hat{E}_\pp 
  + \hat{\omega}\Big(\hat{E}_\pp^2{+}\textstyle{\frac{1}{3}}\hat{\bold{p}}^2\Big)
  + \Big(\hat{\beta}_{\langle\mu\rangle}{-}2\hat{E}_\pp\,\hat{\omega}_{\langle\mu\rangle}\Big)
             \hat{p}^{\langle\mu\rangle}
  + \hat{\omega}_{\langle\mu\nu\rangle}\hat{p}^{\langle\mu}\hat{p}^{\nu\rangle} \right] f_a \,.
%\end{split}
\ee
Here we used that $\hat{\omega}_{\mu\nu}$ is traceless and introduced $\hat{\beta}\equiv-\hat{\beta}_\mu\hat{u}^\mu$, $\hat{\bold{p}}^2\equiv\po/\cosh^2\rho{+}\pp_\eta^2$, $\hat{\omega}\equiv\hat{\omega}^{\mu\nu}\hat{u}_\mu\hat{u}_\nu$, and $\hat{\omega}_{\langle\mu\rangle}= - \hat{\Delta}_{\mu\nu}\hat{\omega}^{\nu\lambda}\hat{u}_\lambda$. This expression can be simplified by using the constraints imposed by Gubser symmetry and Landau matching. In the Landau frame and in the absence of chemical potentials, the spatial vectors $\hat{\beta}_{\langle\mu\rangle}$ and $\hat{\omega}_{\langle\mu\rangle}$ vanish, respectively. Due to the $\text{SO}(3)_q$ symmetry, the traceless symmetric spatial tensor $\hat{\omega}_{\langle\mu\nu\rangle}$ has only one independent component, namely $\hat{\omega}_{\langle\eta\eta\rangle}$:
\be
\hat{\omega}_{\langle\mu\nu\rangle}=\hat{\omega}_{\langle\eta\eta\rangle}\,\text{diag}\left(0,-\cosh^2\rho,-\cosh^2\rho\sin^2\theta,1\right)\,.
\ee
Using the mass-shell condition to eliminate $\po$, Eq.~\eqref{eq:14grad} thus reduces to
\be
\label{eq:14momentC}
\tdf_{14} = \left[\hat{\alpha} - \hat{\beta}\hat{E}_\pp 
              + \hat{\omega}\left(\hat{E}_\pp^2{+}\textstyle{\frac{1}{3}}\hat{\bold{p}}^2\right)  
              + \textstyle{\frac{1}{2}}\hat{\omega}_{\langle\eta\eta\rangle}\left(3\pp_\eta^2{-}\hat{\bold{p}}^2 \right) \right] f_a\,.
\ee
The four coefficients $\hat{\alpha}$, $\hat{\beta}$, $\hat{\omega}$, and $\hat{\omega}_{\langle\eta\eta\rangle}$ must be matched to the contributions of the residual deviation $\tdf$ to the particle and energy densities as well as the bulk and shear viscous pressures \cite{Bazow:2013ifa}:
\be
\label{eq:match4}
\delta\hat{\tilde{n}} \equiv -\langle (\hat{u}\cdot\pp) \rangle_{\tilde{\delta}} = 0,\quad
\delta\hat{\tilde{\epsilon}} \equiv \langle (\hat{u}\cdot\pp)^2 \rangle_{\tilde{\delta}} = 0,\quad
\hat{\tilde{\Pi}} \equiv \frac{1}{3} \langle \hat{\Delta}_{\mu\nu} \pp^\mu \pp^\nu \rangle_{\tilde{\delta}} 
     = \frac{1}{3} \langle \hat{\bold{p}}^2 \rangle_{\tilde{\delta}},\quad
\hat{\tilde{\pi}} \equiv \langle\pp^{\langle\eta}\pp^{\eta\rangle}\rangle_{\tilde{\delta}} 
     = \langle \hat{p}^2_\eta - \frac{1}{3} \hat{\bold{p}}^2 \rangle_{\tilde{\delta}}.
\ee
For conformal systems the bulk viscous pressure vanishes; for technical reasons we introduce an infinitesimal fictitious mass $m$ for the particles, approaching the conformal limit at the end by setting $m\to0$. This is the reason why in Eq.~\eqref{eq:14momentC} we did not replace $\hat{\bold{p}}^2$ by $\hat{E}_\pp^2$.

Inserting the ansatz \eqref{eq:14momentC} into Eq.~\eqref{eq:match4} and using $\hat{\bold{p}}^2 = \hat{E}_\pp^2{-}m^2$ one obtains the matrix equation $\mathcal{A}\,b = c$ where 
\be
b = \left(
\begin{array}{c c c c}
\hat{\alpha} & \hat{\beta} & \hat{\omega} & \hat{\omega}_{\langle\eta\eta\rangle}
\end{array} \right)^T,
\qquad
c = \big(
\begin{array}{c c c c}
0 & 0 & 3\hat{\tilde{\Pi}} & \hat{\tilde{\pi}}
\end{array} \big)^T,
\ee
\be
\mathcal{A} =
\left(
\begin{array}{c c c c}
\\
\Ih_{100}  & \ -\Ih_{200} & \ \frac{4}{3} \Ih_{300} & \ \frac{3}{2}\Ih_{320} - \frac{1}{2}\Ih_{300}  \\ \\
%%%%%
\Ih_{200}  & \ -\Ih_{300} & \ \frac{4}{3} \Ih_{400}  - \frac{1}{3}m^2\Ih_{200} & \ \frac{3}{2}\Ih_{420} - \frac{1}{2}\Ih_{400} + \frac{1}{2}m^2\Ih_{200}\\ \\
%%%%%
\Ih_{200} - m^2\Ih_{000}  & \ -\Ih_{300} + m^2\Ih_{100} & \ \frac{4}{3} \Ih_{400}  - \frac{5}{3}m^2\Ih_{200} & \ \frac{3}{2}\Ih_{420} - \frac{1}{2}\Ih_{400} - \frac{3}{2}m^2\Ih_{220} + m^2\Ih_{200} \\ \\
%%%%%
\Ih_{220} - \frac{1}{3} \Ih_{200} & \ -\Ih_{320} + \frac{1}{3} \Ih_{300} & \ \frac{4}{3} \Ih_{420} - \frac{4}{9}\Ih_{400} & \ \frac{3}{2}\Ih_{440} - \Ih_{420} + \frac{1}{6}\Ih_{400} \\ \\
\end{array} \right).
\ee
Here we expanded the $2^{\text{nd}}$ and $3^{\text{rd}}$ rows in $m$, keeping only terms up to $\mathcal{O}(m^2)$. After subtracting the $2^{\text{nd}}$ from the $3^{\text{rd}}$ row and rescaling it by $-1/m^2$, the matrix equation becomes
\be
\label{eq:matrixsimple}
\left(
\begin{array}{c c c c}
\\
\Ih_{100}  & \ -\Ih_{200} & \ \frac{4}{3} \Ih_{300} & \ \frac{3}{2}\Ih_{320} - \frac{1}{2}\Ih_{300}  \\ \\
%%%%%
\Ih_{200}  & \ -\Ih_{300} & \ \frac{4}{3} \Ih_{400}  - \frac{1}{3}m^2\Ih_{200} & \ \frac{3}{2}\Ih_{420} - \frac{1}{2}\Ih_{400} + \frac{1}{2}m^2\Ih_{200}\\ \\
%%%%%
\Ih_{000}  & \ - \Ih_{100} & \ \frac{4}{3}\Ih_{200} & \ \frac{3}{2}\Ih_{220} - \frac{1}{2}\Ih_{200} \\ \\
%%%%%
\Ih_{220} - \frac{1}{3} \Ih_{200} & \ -\Ih_{320} + \frac{1}{3} \Ih_{300} & \ \frac{4}{3} \Ih_{420} - \frac{4}{9}\Ih_{400} & \ \frac{3}{2}\Ih_{440} - \Ih_{420} + \frac{1}{6}\Ih_{400} \\ \\
\end{array} \right)
\left(
\begin{array}{c}
\\
\hat{\alpha} \\ \\
\hat{\beta} \\ \\
\hat{\omega} \\ \\
\hat{\omega}_{\langle\eta\eta\rangle} \\ \\
\end{array} \right) =
\left(
\begin{array}{c}
\\
0 \\ \\
0 \\ \\ 
\hat{\tilde{\rho}}_0 \\ \\
\hat{\tilde{\pi}} \\ \\
\end{array} \right),
\ee
where $\hat{\tilde{\rho}}_0 \equiv -3\hat{\tilde{\Pi}}/m^2$ is the irreducible $\tdf$-moment associated with the residual bulk viscous pressure. In the conformal limit we set in Eq.~\eqref{eq:matrixsimple} the mass $m$ and the scalar moment $\hat{\tilde{\rho}}_0$ to zero. The matrix can now be easily inverted. One finds
\be
\label{eq:14gradgubcoeff}
\hat{\alpha} = \frac{\hat{\tilde{\pi}} \ \alpha_{\tilde{\pi}}(\xi)}{\J_{2}(\hat{\Lambda})}\,,\qquad
\hat{\beta} = \frac{\hat{\tilde{\pi}} \ \beta_{\tilde{\pi}}(\xi)}{\J_{3}(\hat{\Lambda})} \,,\qquad
\hat{\omega} = \frac{\hat{\tilde{\pi}} \ \gamma_{\tilde{\pi}}(\xi)}{\J_{4}(\hat{\Lambda})} \,,\qquad
\hat{\omega}_{\langle\eta\eta\rangle} = \frac{\hat{\tilde{\pi}} \ \kappa_{\tilde{\pi}}(\xi)}{\J_{4}(\hat{\Lambda})},
\ee
where, with the shorthand $t(\xi)\equiv\arctan\sqrt{\xi}/\sqrt{\xi}$, the functions $\alpha_{\tilde{\pi}}$, $\beta_{\tilde{\pi}}$, $\gamma_{\tilde{\pi}}$ and $\kappa_{\tilde{\pi}}$ are given by
\be 
\label{eq:alphapi}
\begin{gathered}
%\textstyle
\alpha_{\tilde{\pi}}(\xi) = -3\xi\,
 \frac{2\big(39{+}76\xi\big){\,-\,}\big(213 + 343 \xi + 104 \xi^2\big) t(\xi)%
                {\,+\,}90\big(1{+}\xi\big)t^2(\xi){+\,}45\big(1{+}\xi\big)^3 t^3(\xi)}
        {192 + 6\big(131{+}212\xi\big)t(\xi){\,-\,}\big(1932{+}2817\xi{+}712 \xi^2\big)t^2(\xi)%
             {\,+\,}630\big(1{+}\xi\big)t^3(\xi){\,+\,}315\big(1{+}\xi\big)^3t^4(\xi)},
\end{gathered}
\ee
\be 
\label{eq:betapi}
\begin{gathered}
%\textstyle
\beta_{\tilde{\pi}}(\xi) = 24\xi\sqrt{1{+}\xi}\,
 \frac{12{\,+\,}\big(15{+}13\xi\big)t(\xi){\,-\,}\big(27{+}26\xi{-}\xi^2\big)t^2(\xi)}
        {192 + 6\big(131{+}212\xi\big)t(\xi){\,-\,}\big(1932{+}2817\xi{+}712 \xi^2\big)t^2(\xi)%
             {\,+\,}630\big(1{+}\xi\big)t^3(\xi){\,+\,}315\big(1{+}\xi\big)^3t^4(\xi)},
\end{gathered}
\ee
\be 
\label{eq:gammapi}
\begin{gathered}
%\textstyle
\gamma_{\tilde{\pi}}(\xi) = \frac{5}{2}\xi\,
             \frac{3\big(57{+}95\xi{+}4\xi^2\big){\,-\,}2\big(99{+}234\xi{+}231\xi^2{+}122\xi^3\big)t(\xi)%
                    {\,+\,}9\big(9{+}19\xi{+}31\xi^2{+}21\xi^3\big)t^2(\xi){\,+\,}54\big(1{+}\xi\big)^3(2\xi{-}1)t^3(\xi)}
       {192{\,+\,}6\big(212\xi{+}131\big) t(\xi){\,-\,}\big(1923{+}2817\xi{+}712\xi^2\big) t^2(\xi)%
               {\,+\,}630\big(1{+}\xi\big)t^3(\xi){\,+\,}315\big(1{+}\xi\big)^3 t^4(\xi)},
\end{gathered}
\ee
\be
\label{eq:kappapi}
\begin{gathered}
%\textstyle
\kappa_{\tilde{\pi}}(\xi) = \frac{20}{3}\xi^2\,
            \frac{3(4\xi{-}13){\,-\,}\big(162{+}297\xi{+}122\xi^2\big)t(\xi)%
                   {\,+\,}9\big(31{+}52\xi{+}21\xi^2\big)t^2(\xi){\,+\,}108\big(1{+}\xi\big)^3 t^3(\xi)}
        {192{\,+\,}6\big(212\xi{+}131\big) t(\xi){\,-\,}\big(1923{+}2817\xi{+}712\xi^2\big) t^2(\xi)%
               {\,+\,}630\big(1{+}\xi\big)t^3(\xi){\,+\,}315\big(1{+}\xi\big)^3 t^4(\xi)}.
\end{gathered}
\ee
\end{widetext}%
In the massless limit Eq.~\eqref{eq:14momentC} coincides with Eq.~\eqref{eq:14gradgub} in the text. As a cross-check we consider the isotropic limit $\xi \rightarrow 0$ of the coefficients \eqref{eq:14gradgubcoeff}: 
\bea
&&\lim_{\xi\to 0}\hat{\alpha} = 0,\quad 
\lim_{\xi\to 0}\hat{\beta}= 0,
\nonumber\\
&& \lim_{\xi\to 0}\hat{\omega} = 0,\quad
\lim_{\xi\to 0}\hat{\omega}_{\langle\eta\eta\rangle} 
     = \frac{\hat{\pi}}{2(\hat{\epsilon}+\hat{\mathcal{P}})\hat{T}^2}.
\eea
With this we see that Eq.~\eqref{eq:14gradgub} reduces in the isotropic limit to the well-known Israel-Stewart result \cite{Israel:1979wp} for conformal systems \cite{Baier:2007ix}:
\be
\label{eq:ISlimit}
\lim_{\xi\to 0}\delta \tilde{f}_{14} = 
\left[\frac{\hat{\pi}}{4(\hat{\epsilon}{+}\hat{\mathcal{P}})\hat{T}^2}\left(3\pvv^2 - \hat{E}_\pp^2 \right) \right] \feq.
\ee
%

%%%%%%%%%%%%%%% References %%%%%%%%%%%%%%%%%%%%%%%%%%%%%%%%%

\bibliography{ahydrogubser}

\begin{thebibliography}{67}%
\makeatletter
\providecommand \@ifxundefined [1]{%
 \@ifx{#1\undefined}
}%
\providecommand \@ifnum [1]{%
 \ifnum #1\expandafter \@firstoftwo
 \else \expandafter \@secondoftwo
 \fi
}%
\providecommand \@ifx [1]{%
 \ifx #1\expandafter \@firstoftwo
 \else \expandafter \@secondoftwo
 \fi
}%
\providecommand \natexlab [1]{#1}%
\providecommand \enquote  [1]{``#1''}%
\providecommand \bibnamefont  [1]{#1}%
\providecommand \bibfnamefont [1]{#1}%
\providecommand \citenamefont [1]{#1}%
\providecommand \href@noop [0]{\@secondoftwo}%
\providecommand \href [0]{\begingroup \@sanitize@url \@href}%
\providecommand \@href[1]{\@@startlink{#1}\@@href}%
\providecommand \@@href[1]{\endgroup#1\@@endlink}%
\providecommand \@sanitize@url [0]{\catcode `\\12\catcode `\$12\catcode
  `\&12\catcode `\#12\catcode `\^12\catcode `\_12\catcode `\%12\relax}%
\providecommand \@@startlink[1]{}%
\providecommand \@@endlink[0]{}%
\providecommand \url  [0]{\begingroup\@sanitize@url \@url }%
\providecommand \@url [1]{\endgroup\@href {#1}{\urlprefix }}%
\providecommand \urlprefix  [0]{URL }%
\providecommand \Eprint [0]{\href }%
\providecommand \doibase [0]{http://dx.doi.org/}%
\providecommand \selectlanguage [0]{\@gobble}%
\providecommand \bibinfo  [0]{\@secondoftwo}%
\providecommand \bibfield  [0]{\@secondoftwo}%
\providecommand \translation [1]{[#1]}%
\providecommand \BibitemOpen [0]{}%
\providecommand \bibitemStop [0]{}%
\providecommand \bibitemNoStop [0]{.\EOS\space}%
\providecommand \EOS [0]{\spacefactor3000\relax}%
\providecommand \BibitemShut  [1]{\csname bibitem#1\endcsname}%
\let\auto@bib@innerbib\@empty
%</preamble>
\bibitem [{\citenamefont {Martinez}\ and\ \citenamefont
  {Strickland}(2010)}]{Martinez:2010sc}%
  \BibitemOpen
  \bibfield  {author} {\bibinfo {author} {\bibfnamefont {M.}~\bibnamefont
  {Martinez}}\ and\ \bibinfo {author} {\bibfnamefont {M.}~\bibnamefont
  {Strickland}},\ }\href {\doibase 10.1016/j.nuclphysa.2010.08.011} {\bibfield
  {journal} {\bibinfo  {journal} {Nucl. Phys.}\ }\textbf {\bibinfo {volume}
  {A848}},\ \bibinfo {pages} {183} (\bibinfo {year} {2010})},\ \Eprint
  {http://arxiv.org/abs/1007.0889} {arXiv:1007.0889 [nucl-th]} \BibitemShut
  {NoStop}%
%%CITATION = 1007.0889;%%
\bibitem [{\citenamefont {Martinez}\ and\ \citenamefont
  {Strickland}(2011)}]{Martinez:2010sd}%
  \BibitemOpen
  \bibfield  {author} {\bibinfo {author} {\bibfnamefont {M.}~\bibnamefont
  {Martinez}}\ and\ \bibinfo {author} {\bibfnamefont {M.}~\bibnamefont
  {Strickland}},\ }\href {\doibase 10.1016/j.nuclphysa.2011.02.003} {\bibfield
  {journal} {\bibinfo  {journal} {Nucl. Phys.}\ }\textbf {\bibinfo {volume}
  {A856}},\ \bibinfo {pages} {68} (\bibinfo {year} {2011})},\ \Eprint
  {http://arxiv.org/abs/1011.3056} {arXiv:1011.3056 [nucl-th]} \BibitemShut
  {NoStop}%
%%CITATION = ARXIV:1011.3056;%%
\bibitem [{\citenamefont {Florkowski}\ and\ \citenamefont
  {Ryblewski}(2011)}]{Florkowski:2010cf}%
  \BibitemOpen
  \bibfield  {author} {\bibinfo {author} {\bibfnamefont {W.}~\bibnamefont
  {Florkowski}}\ and\ \bibinfo {author} {\bibfnamefont {R.}~\bibnamefont
  {Ryblewski}},\ }\href {\doibase 10.1103/PhysRevC.83.034907} {\bibfield
  {journal} {\bibinfo  {journal} {Phys. Rev.}\ }\textbf {\bibinfo {volume}
  {C83}},\ \bibinfo {pages} {034907} (\bibinfo {year} {2011})},\ \Eprint
  {http://arxiv.org/abs/1007.0130} {arXiv:1007.0130 [nucl-th]} \BibitemShut
  {NoStop}%
%%CITATION = ARXIV:1007.0130;%%
\bibitem [{\citenamefont {Ryblewski}\ and\ \citenamefont
  {Florkowski}(2011{\natexlab{a}})}]{Ryblewski:2010bs}%
  \BibitemOpen
  \bibfield  {author} {\bibinfo {author} {\bibfnamefont {R.}~\bibnamefont
  {Ryblewski}}\ and\ \bibinfo {author} {\bibfnamefont {W.}~\bibnamefont
  {Florkowski}},\ }\href {\doibase 10.1088/0954-3899/38/1/015104} {\bibfield
  {journal} {\bibinfo  {journal} {J. Phys.}\ }\textbf {\bibinfo {volume}
  {G38}},\ \bibinfo {pages} {015104} (\bibinfo {year} {2011}{\natexlab{a}})},\
  \Eprint {http://arxiv.org/abs/1007.4662} {arXiv:1007.4662 [nucl-th]}
  \BibitemShut {NoStop}%
%%CITATION = ARXIV:1007.4662;%%
\bibitem [{\citenamefont {Florkowski}\ and\ \citenamefont
  {Ryblewski}(2012)}]{Florkowski:2011jg}%
  \BibitemOpen
  \bibfield  {author} {\bibinfo {author} {\bibfnamefont {W.}~\bibnamefont
  {Florkowski}}\ and\ \bibinfo {author} {\bibfnamefont {R.}~\bibnamefont
  {Ryblewski}},\ }\href {\doibase 10.1103/PhysRevC.85.044902} {\bibfield
  {journal} {\bibinfo  {journal} {Phys. Rev.}\ }\textbf {\bibinfo {volume}
  {C85}},\ \bibinfo {pages} {044902} (\bibinfo {year} {2012})},\ \Eprint
  {http://arxiv.org/abs/1111.5997} {arXiv:1111.5997 [nucl-th]} \BibitemShut
  {NoStop}%
%%CITATION = ARXIV:1111.5997;%%
\bibitem [{\citenamefont {Ryblewski}\ and\ \citenamefont
  {Florkowski}(2011{\natexlab{b}})}]{Ryblewski:2011aq}%
  \BibitemOpen
  \bibfield  {author} {\bibinfo {author} {\bibfnamefont {R.}~\bibnamefont
  {Ryblewski}}\ and\ \bibinfo {author} {\bibfnamefont {W.}~\bibnamefont
  {Florkowski}},\ }\href {\doibase 10.1140/epjc/s10052-011-1761-8} {\bibfield
  {journal} {\bibinfo  {journal} {Eur. Phys. J.}\ }\textbf {\bibinfo {volume}
  {C71}},\ \bibinfo {pages} {1761} (\bibinfo {year} {2011}{\natexlab{b}})},\
  \Eprint {http://arxiv.org/abs/1103.1260} {arXiv:1103.1260 [nucl-th]}
  \BibitemShut {NoStop}%
%%CITATION = ARXIV:1103.1260;%%
\bibitem [{\citenamefont {Martinez}\ \emph {et~al.}(2012)\citenamefont
  {Martinez}, \citenamefont {Ryblewski},\ and\ \citenamefont
  {Strickland}}]{Martinez:2012tu}%
  \BibitemOpen
  \bibfield  {author} {\bibinfo {author} {\bibfnamefont {M.}~\bibnamefont
  {Martinez}}, \bibinfo {author} {\bibfnamefont {R.}~\bibnamefont {Ryblewski}},
  \ and\ \bibinfo {author} {\bibfnamefont {M.}~\bibnamefont {Strickland}},\
  }\href {\doibase 10.1103/PhysRevC.85.064913} {\bibfield  {journal} {\bibinfo
  {journal} {Phys. Rev.}\ }\textbf {\bibinfo {volume} {C85}},\ \bibinfo {pages}
  {064913} (\bibinfo {year} {2012})},\ \Eprint {http://arxiv.org/abs/1204.1473}
  {arXiv:1204.1473 [nucl-th]} \BibitemShut {NoStop}%
%%CITATION = ARXIV:1204.1473;%%
\bibitem [{\citenamefont {Ryblewski}\ and\ \citenamefont
  {Florkowski}(2012)}]{Ryblewski:2012rr}%
  \BibitemOpen
  \bibfield  {author} {\bibinfo {author} {\bibfnamefont {R.}~\bibnamefont
  {Ryblewski}}\ and\ \bibinfo {author} {\bibfnamefont {W.}~\bibnamefont
  {Florkowski}},\ }\href {\doibase 10.1103/PhysRevC.85.064901} {\bibfield
  {journal} {\bibinfo  {journal} {Phys. Rev.}\ }\textbf {\bibinfo {volume}
  {C85}},\ \bibinfo {pages} {064901} (\bibinfo {year} {2012})},\ \Eprint
  {http://arxiv.org/abs/1204.2624} {arXiv:1204.2624 [nucl-th]} \BibitemShut
  {NoStop}%
%%CITATION = ARXIV:1204.2624;%%
\bibitem [{\citenamefont {Florkowski}\ \emph
  {et~al.}(2013{\natexlab{a}})\citenamefont {Florkowski}, \citenamefont {Maj},
  \citenamefont {Ryblewski},\ and\ \citenamefont
  {Strickland}}]{Florkowski:2012as}%
  \BibitemOpen
  \bibfield  {author} {\bibinfo {author} {\bibfnamefont {W.}~\bibnamefont
  {Florkowski}}, \bibinfo {author} {\bibfnamefont {R.}~\bibnamefont {Maj}},
  \bibinfo {author} {\bibfnamefont {R.}~\bibnamefont {Ryblewski}}, \ and\
  \bibinfo {author} {\bibfnamefont {M.}~\bibnamefont {Strickland}},\ }\href
  {\doibase 10.1103/PhysRevC.87.034914} {\bibfield  {journal} {\bibinfo
  {journal} {Phys. Rev.}\ }\textbf {\bibinfo {volume} {C87}},\ \bibinfo {pages}
  {034914} (\bibinfo {year} {2013}{\natexlab{a}})},\ \Eprint
  {http://arxiv.org/abs/1209.3671} {arXiv:1209.3671 [nucl-th]} \BibitemShut
  {NoStop}%
%%CITATION = ARXIV:1209.3671;%%
\bibitem [{\citenamefont {Florkowski}\ and\ \citenamefont
  {Maj}(2013)}]{Florkowski:2013uqa}%
  \BibitemOpen
  \bibfield  {author} {\bibinfo {author} {\bibfnamefont {W.}~\bibnamefont
  {Florkowski}}\ and\ \bibinfo {author} {\bibfnamefont {R.}~\bibnamefont
  {Maj}},\ }\href {\doibase 10.5506/APhysPolB.44.2003} {\bibfield  {journal}
  {\bibinfo  {journal} {Acta Phys. Polon.}\ }\textbf {\bibinfo {volume}
  {B44}},\ \bibinfo {pages} {2003} (\bibinfo {year} {2013})},\ \Eprint
  {http://arxiv.org/abs/1309.2786} {arXiv:1309.2786 [nucl-th]} \BibitemShut
  {NoStop}%
%%CITATION = ARXIV:1309.2786;%%
\bibitem [{\citenamefont {Bazow}\ \emph {et~al.}(2014)\citenamefont {Bazow},
  \citenamefont {Heinz},\ and\ \citenamefont {Strickland}}]{Bazow:2013ifa}%
  \BibitemOpen
  \bibfield  {author} {\bibinfo {author} {\bibfnamefont {D.}~\bibnamefont
  {Bazow}}, \bibinfo {author} {\bibfnamefont {U.}~\bibnamefont {Heinz}}, \ and\
  \bibinfo {author} {\bibfnamefont {M.}~\bibnamefont {Strickland}},\ }\href
  {\doibase 10.1103/PhysRevC.90.054910} {\bibfield  {journal} {\bibinfo
  {journal} {Phys. Rev.}\ }\textbf {\bibinfo {volume} {C90}},\ \bibinfo {pages}
  {054910} (\bibinfo {year} {2014})},\ \Eprint {http://arxiv.org/abs/1311.6720}
  {arXiv:1311.6720 [nucl-th]} \BibitemShut {NoStop}%
%%CITATION = ARXIV:1311.6720;%%
\bibitem [{\citenamefont {Tinti}\ and\ \citenamefont
  {Florkowski}(2014)}]{Tinti:2013vba}%
  \BibitemOpen
  \bibfield  {author} {\bibinfo {author} {\bibfnamefont {L.}~\bibnamefont
  {Tinti}}\ and\ \bibinfo {author} {\bibfnamefont {W.}~\bibnamefont
  {Florkowski}},\ }\href {\doibase 10.1103/PhysRevC.89.034907} {\bibfield
  {journal} {\bibinfo  {journal} {Phys. Rev.}\ }\textbf {\bibinfo {volume}
  {C89}},\ \bibinfo {pages} {034907} (\bibinfo {year} {2014})},\ \Eprint
  {http://arxiv.org/abs/1312.6614} {arXiv:1312.6614 [nucl-th]} \BibitemShut
  {NoStop}%
%%CITATION = ARXIV:1312.6614;%%
\bibitem [{\citenamefont {Florkowski}\ \emph
  {et~al.}(2014{\natexlab{a}})\citenamefont {Florkowski}, \citenamefont
  {Ryblewski}, \citenamefont {Strickland},\ and\ \citenamefont
  {Tinti}}]{Florkowski:2014bba}%
  \BibitemOpen
  \bibfield  {author} {\bibinfo {author} {\bibfnamefont {W.}~\bibnamefont
  {Florkowski}}, \bibinfo {author} {\bibfnamefont {R.}~\bibnamefont
  {Ryblewski}}, \bibinfo {author} {\bibfnamefont {M.}~\bibnamefont
  {Strickland}}, \ and\ \bibinfo {author} {\bibfnamefont {L.}~\bibnamefont
  {Tinti}},\ }\href {\doibase 10.1103/PhysRevC.89.054909} {\bibfield  {journal}
  {\bibinfo  {journal} {Phys. Rev.}\ }\textbf {\bibinfo {volume} {C89}},\
  \bibinfo {pages} {054909} (\bibinfo {year} {2014}{\natexlab{a}})},\ \Eprint
  {http://arxiv.org/abs/1403.1223} {arXiv:1403.1223 [hep-ph]} \BibitemShut
  {NoStop}%
%%CITATION = ARXIV:1403.1223;%%
\bibitem [{\citenamefont {Florkowski}\ and\ \citenamefont
  {Madetko}(2014)}]{Florkowski:2014txa}%
  \BibitemOpen
  \bibfield  {author} {\bibinfo {author} {\bibfnamefont {W.}~\bibnamefont
  {Florkowski}}\ and\ \bibinfo {author} {\bibfnamefont {O.}~\bibnamefont
  {Madetko}},\ }\href {\doibase 10.5506/APhysPolB.45.1103} {\bibfield
  {journal} {\bibinfo  {journal} {Acta Phys. Polon.}\ }\textbf {\bibinfo
  {volume} {B45}},\ \bibinfo {pages} {1103} (\bibinfo {year} {2014})},\ \Eprint
  {http://arxiv.org/abs/1402.2401} {arXiv:1402.2401 [nucl-th]} \BibitemShut
  {NoStop}%
%%CITATION = ARXIV:1402.2401;%%
\bibitem [{\citenamefont {Nopoush}\ \emph {et~al.}(2014)\citenamefont
  {Nopoush}, \citenamefont {Ryblewski},\ and\ \citenamefont
  {Strickland}}]{Nopoush:2014pfa}%
  \BibitemOpen
  \bibfield  {author} {\bibinfo {author} {\bibfnamefont {M.}~\bibnamefont
  {Nopoush}}, \bibinfo {author} {\bibfnamefont {R.}~\bibnamefont {Ryblewski}},
  \ and\ \bibinfo {author} {\bibfnamefont {M.}~\bibnamefont {Strickland}},\
  }\href {\doibase 10.1103/PhysRevC.90.014908} {\bibfield  {journal} {\bibinfo
  {journal} {Phys. Rev.}\ }\textbf {\bibinfo {volume} {C90}},\ \bibinfo {pages}
  {014908} (\bibinfo {year} {2014})},\ \Eprint {http://arxiv.org/abs/1405.1355}
  {arXiv:1405.1355 [hep-ph]} \BibitemShut {NoStop}%
%%CITATION = ARXIV:1405.1355;%%
\bibitem [{\citenamefont {Denicol}\ \emph
  {et~al.}(2014{\natexlab{a}})\citenamefont {Denicol}, \citenamefont
  {Florkowski}, \citenamefont {Ryblewski},\ and\ \citenamefont
  {Strickland}}]{Denicol:2014mca}%
  \BibitemOpen
  \bibfield  {author} {\bibinfo {author} {\bibfnamefont {G.~S.}\ \bibnamefont
  {Denicol}}, \bibinfo {author} {\bibfnamefont {W.}~\bibnamefont {Florkowski}},
  \bibinfo {author} {\bibfnamefont {R.}~\bibnamefont {Ryblewski}}, \ and\
  \bibinfo {author} {\bibfnamefont {M.}~\bibnamefont {Strickland}},\ }\href
  {\doibase 10.1103/PhysRevC.90.044905} {\bibfield  {journal} {\bibinfo
  {journal} {Phys. Rev.}\ }\textbf {\bibinfo {volume} {C90}},\ \bibinfo {pages}
  {044905} (\bibinfo {year} {2014}{\natexlab{a}})},\ \Eprint
  {http://arxiv.org/abs/1407.4767} {arXiv:1407.4767 [hep-ph]} \BibitemShut
  {NoStop}%
%%CITATION = ARXIV:1407.4767;%%
\bibitem [{\citenamefont {Bazow}\ \emph {et~al.}(2015)\citenamefont {Bazow},
  \citenamefont {Heinz},\ and\ \citenamefont {Martinez}}]{Bazow:2015cha}%
  \BibitemOpen
  \bibfield  {author} {\bibinfo {author} {\bibfnamefont {D.}~\bibnamefont
  {Bazow}}, \bibinfo {author} {\bibfnamefont {U.}~\bibnamefont {Heinz}}, \ and\
  \bibinfo {author} {\bibfnamefont {M.}~\bibnamefont {Martinez}},\ }\href
  {\doibase 10.1103/PhysRevC.91.064903} {\bibfield  {journal} {\bibinfo
  {journal} {Phys. Rev.}\ }\textbf {\bibinfo {volume} {C91}},\ \bibinfo {pages}
  {064903} (\bibinfo {year} {2015})},\ \Eprint
  {http://arxiv.org/abs/1503.07443} {arXiv:1503.07443 [nucl-th]} \BibitemShut
  {NoStop}%
%%CITATION = ARXIV:1503.07443;%%
\bibitem [{\citenamefont {Nopoush}\ \emph
  {et~al.}(2015{\natexlab{a}})\citenamefont {Nopoush}, \citenamefont
  {Strickland}, \citenamefont {Ryblewski}, \citenamefont {Bazow}, \citenamefont
  {Heinz},\ and\ \citenamefont {Martinez}}]{Nopoush:2015yga}%
  \BibitemOpen
  \bibfield  {author} {\bibinfo {author} {\bibfnamefont {M.}~\bibnamefont
  {Nopoush}}, \bibinfo {author} {\bibfnamefont {M.}~\bibnamefont {Strickland}},
  \bibinfo {author} {\bibfnamefont {R.}~\bibnamefont {Ryblewski}}, \bibinfo
  {author} {\bibfnamefont {D.}~\bibnamefont {Bazow}}, \bibinfo {author}
  {\bibfnamefont {U.}~\bibnamefont {Heinz}}, \ and\ \bibinfo {author}
  {\bibfnamefont {M.}~\bibnamefont {Martinez}},\ }\href {\doibase
  10.1103/PhysRevC.92.044912} {\bibfield  {journal} {\bibinfo  {journal} {Phys.
  Rev.}\ }\textbf {\bibinfo {volume} {C92}},\ \bibinfo {pages} {044912}
  (\bibinfo {year} {2015}{\natexlab{a}})},\ \Eprint
  {http://arxiv.org/abs/1506.05278} {arXiv:1506.05278 [nucl-th]} \BibitemShut
  {NoStop}%
%%CITATION = ARXIV:1506.05278;%%
\bibitem [{\citenamefont {Florkowski}\ \emph {et~al.}(2015)\citenamefont
  {Florkowski}, \citenamefont {Maksymiuk}, \citenamefont {Ryblewski},\ and\
  \citenamefont {Tinti}}]{Florkowski:2015cba}%
  \BibitemOpen
  \bibfield  {author} {\bibinfo {author} {\bibfnamefont {W.}~\bibnamefont
  {Florkowski}}, \bibinfo {author} {\bibfnamefont {E.}~\bibnamefont
  {Maksymiuk}}, \bibinfo {author} {\bibfnamefont {R.}~\bibnamefont
  {Ryblewski}}, \ and\ \bibinfo {author} {\bibfnamefont {L.}~\bibnamefont
  {Tinti}},\ }\href {\doibase 10.1103/PhysRevC.92.054912} {\bibfield  {journal}
  {\bibinfo  {journal} {Phys. Rev.}\ }\textbf {\bibinfo {volume} {C92}},\
  \bibinfo {pages} {054912} (\bibinfo {year} {2015})},\ \Eprint
  {http://arxiv.org/abs/1508.04534} {arXiv:1508.04534 [nucl-th]} \BibitemShut
  {NoStop}%
%%CITATION = ARXIV:1508.04534;%%
\bibitem [{\citenamefont {Alqahtani}\ \emph {et~al.}(2015)\citenamefont
  {Alqahtani}, \citenamefont {Nopoush},\ and\ \citenamefont
  {Strickland}}]{Alqahtani:2015qja}%
  \BibitemOpen
  \bibfield  {author} {\bibinfo {author} {\bibfnamefont {M.}~\bibnamefont
  {Alqahtani}}, \bibinfo {author} {\bibfnamefont {M.}~\bibnamefont {Nopoush}},
  \ and\ \bibinfo {author} {\bibfnamefont {M.}~\bibnamefont {Strickland}},\
  }\href {\doibase 10.1103/PhysRevC.92.054910} {\bibfield  {journal} {\bibinfo
  {journal} {Phys. Rev.}\ }\textbf {\bibinfo {volume} {C92}},\ \bibinfo {pages}
  {054910} (\bibinfo {year} {2015})},\ \Eprint
  {http://arxiv.org/abs/1509.02913} {arXiv:1509.02913 [hep-ph]} \BibitemShut
  {NoStop}%
%%CITATION = ARXIV:1509.02913;%%
\bibitem [{\citenamefont {Bazow}\ \emph
  {et~al.}(2016{\natexlab{a}})\citenamefont {Bazow}, \citenamefont {Martinez},\
  and\ \citenamefont {Heinz}}]{Bazow:2015zca}%
  \BibitemOpen
  \bibfield  {author} {\bibinfo {author} {\bibfnamefont {D.}~\bibnamefont
  {Bazow}}, \bibinfo {author} {\bibfnamefont {M.}~\bibnamefont {Martinez}}, \
  and\ \bibinfo {author} {\bibfnamefont {U.}~\bibnamefont {Heinz}},\ }\href
  {\doibase 10.1103/PhysRevD.93.034002} {\bibfield  {journal} {\bibinfo
  {journal} {Phys. Rev.}\ }\textbf {\bibinfo {volume} {D93}},\ \bibinfo {pages}
  {034002} (\bibinfo {year} {2016}{\natexlab{a}})},\ \Eprint
  {http://arxiv.org/abs/1507.06595} {arXiv:1507.06595 [nucl-th]} \BibitemShut
  {NoStop}%
%%CITATION = ARXIV:1507.06595;%%
\bibitem [{\citenamefont {Tinti}\ \emph {et~al.}(2016)\citenamefont {Tinti},
  \citenamefont {Ryblewski}, \citenamefont {Florkowski},\ and\ \citenamefont
  {Strickland}}]{Tinti:2015xra}%
  \BibitemOpen
  \bibfield  {author} {\bibinfo {author} {\bibfnamefont {L.}~\bibnamefont
  {Tinti}}, \bibinfo {author} {\bibfnamefont {R.}~\bibnamefont {Ryblewski}},
  \bibinfo {author} {\bibfnamefont {W.}~\bibnamefont {Florkowski}}, \ and\
  \bibinfo {author} {\bibfnamefont {M.}~\bibnamefont {Strickland}},\ }\href
  {\doibase 10.1016/j.nuclphysa.2015.11.006} {\bibfield  {journal} {\bibinfo
  {journal} {Nucl. Phys.}\ }\textbf {\bibinfo {volume} {A946}},\ \bibinfo
  {pages} {29} (\bibinfo {year} {2016})},\ \Eprint
  {http://arxiv.org/abs/1505.06456} {arXiv:1505.06456 [hep-ph]} \BibitemShut
  {NoStop}%
%%CITATION = ARXIV:1505.06456;%%
\bibitem [{\citenamefont {Bluhm}\ and\ \citenamefont
  {Schäfer}(2015)}]{Bluhm:2015raa}%
  \BibitemOpen
  \bibfield  {author} {\bibinfo {author} {\bibfnamefont {M.}~\bibnamefont
  {Bluhm}}\ and\ \bibinfo {author} {\bibfnamefont {T.}~\bibnamefont
  {Schäfer}},\ }\href {\doibase 10.1103/PhysRevA.92.043602} {\bibfield
  {journal} {\bibinfo  {journal} {Phys. Rev.}\ }\textbf {\bibinfo {volume}
  {A92}},\ \bibinfo {pages} {043602} (\bibinfo {year} {2015})},\ \Eprint
  {http://arxiv.org/abs/1505.00846} {arXiv:1505.00846 [cond-mat.quant-gas]}
  \BibitemShut {NoStop}%
%%CITATION = ARXIV:1505.00846;%%
\bibitem [{\citenamefont {Bluhm}\ and\ \citenamefont
  {Schaefer}(2016)}]{Bluhm:2015bzi}%
  \BibitemOpen
  \bibfield  {author} {\bibinfo {author} {\bibfnamefont {M.}~\bibnamefont
  {Bluhm}}\ and\ \bibinfo {author} {\bibfnamefont {T.}~\bibnamefont
  {Schaefer}},\ }\href {\doibase 10.1103/PhysRevLett.116.115301} {\bibfield
  {journal} {\bibinfo  {journal} {Phys. Rev. Lett.}\ }\textbf {\bibinfo
  {volume} {116}},\ \bibinfo {pages} {115301} (\bibinfo {year} {2016})},\
  \Eprint {http://arxiv.org/abs/1512.00862} {arXiv:1512.00862
  [cond-mat.quant-gas]} \BibitemShut {NoStop}%
%%CITATION = ARXIV:1512.00862;%%
\bibitem [{\citenamefont {Alqahtani}\ \emph {et~al.}(2016)\citenamefont
  {Alqahtani}, \citenamefont {Nopoush},\ and\ \citenamefont
  {Strickland}}]{Alqahtani:2016rth}%
  \BibitemOpen
  \bibfield  {author} {\bibinfo {author} {\bibfnamefont {M.}~\bibnamefont
  {Alqahtani}}, \bibinfo {author} {\bibfnamefont {M.}~\bibnamefont {Nopoush}},
  \ and\ \bibinfo {author} {\bibfnamefont {M.}~\bibnamefont {Strickland}},\
  }\href@noop {} {\  (\bibinfo {year} {2016})},\ \Eprint
  {http://arxiv.org/abs/1605.02101} {arXiv:1605.02101 [nucl-th]} \BibitemShut
  {NoStop}%
%%CITATION = ARXIV:1605.02101;%%
\bibitem [{\citenamefont {Florkowski}\ \emph
  {et~al.}(2016{\natexlab{a}})\citenamefont {Florkowski}, \citenamefont
  {Ryblewski}, \citenamefont {Strickland},\ and\ \citenamefont
  {Tinti}}]{Florkowski:2016kjj}%
  \BibitemOpen
  \bibfield  {author} {\bibinfo {author} {\bibfnamefont {W.}~\bibnamefont
  {Florkowski}}, \bibinfo {author} {\bibfnamefont {R.}~\bibnamefont
  {Ryblewski}}, \bibinfo {author} {\bibfnamefont {M.}~\bibnamefont
  {Strickland}}, \ and\ \bibinfo {author} {\bibfnamefont {L.}~\bibnamefont
  {Tinti}},\ }\href {\doibase 10.1103/PhysRevC.94.064903} {\bibfield  {journal}
  {\bibinfo  {journal} {Phys. Rev.}\ }\textbf {\bibinfo {volume} {C94}},\
  \bibinfo {pages} {064903} (\bibinfo {year} {2016}{\natexlab{a}})},\ \Eprint
  {http://arxiv.org/abs/1609.06293} {arXiv:1609.06293 [nucl-th]} \BibitemShut
  {NoStop}%
%%CITATION = ARXIV:1609.06293;%%
\bibitem [{\citenamefont {Nopoush}\ \emph {et~al.}(2016)\citenamefont
  {Nopoush}, \citenamefont {Strickland},\ and\ \citenamefont
  {Ryblewski}}]{Nopoush:2016qas}%
  \BibitemOpen
  \bibfield  {author} {\bibinfo {author} {\bibfnamefont {M.}~\bibnamefont
  {Nopoush}}, \bibinfo {author} {\bibfnamefont {M.}~\bibnamefont {Strickland}},
  \ and\ \bibinfo {author} {\bibfnamefont {R.}~\bibnamefont {Ryblewski}},\
  }\href@noop {} {\  (\bibinfo {year} {2016})},\ \Eprint
  {http://arxiv.org/abs/1610.10055} {arXiv:1610.10055 [nucl-th]} \BibitemShut
  {NoStop}%
%%CITATION = ARXIV:1610.10055;%%
\bibitem [{\citenamefont {Florkowski}\ \emph
  {et~al.}(2016{\natexlab{b}})\citenamefont {Florkowski}, \citenamefont
  {Ryblewski},\ and\ \citenamefont {Spaliński}}]{Florkowski:2016zsi}%
  \BibitemOpen
  \bibfield  {author} {\bibinfo {author} {\bibfnamefont {W.}~\bibnamefont
  {Florkowski}}, \bibinfo {author} {\bibfnamefont {R.}~\bibnamefont
  {Ryblewski}}, \ and\ \bibinfo {author} {\bibfnamefont {M.}~\bibnamefont
  {Spaliński}},\ }\href {\doibase 10.1103/PhysRevD.94.114025} {\bibfield
  {journal} {\bibinfo  {journal} {Phys. Rev.}\ }\textbf {\bibinfo {volume}
  {D94}},\ \bibinfo {pages} {114025} (\bibinfo {year} {2016}{\natexlab{b}})},\
  \Eprint {http://arxiv.org/abs/1608.07558} {arXiv:1608.07558 [nucl-th]}
  \BibitemShut {NoStop}%
%%CITATION = ARXIV:1608.07558;%%
\bibitem [{\citenamefont {Schäfer}(2016)}]{Schaefer:2016yzd}%
  \BibitemOpen
  \bibfield  {author} {\bibinfo {author} {\bibfnamefont {T.}~\bibnamefont
  {Schäfer}},\ }\href {\doibase 10.1103/PhysRevA.94.043644} {\bibfield
  {journal} {\bibinfo  {journal} {Phys. Rev.}\ }\textbf {\bibinfo {volume}
  {A94}},\ \bibinfo {pages} {043644} (\bibinfo {year} {2016})},\ \Eprint
  {http://arxiv.org/abs/1608.05083} {arXiv:1608.05083 [cond-mat.quant-gas]}
  \BibitemShut {NoStop}%
%%CITATION = ARXIV:1608.05083;%%
\bibitem [{\citenamefont {Denicol}\ \emph {et~al.}(2012)\citenamefont
  {Denicol}, \citenamefont {Niemi}, \citenamefont {Molnar},\ and\ \citenamefont
  {Rischke}}]{Denicol:2012cn}%
  \BibitemOpen
  \bibfield  {author} {\bibinfo {author} {\bibfnamefont {G.~S.}\ \bibnamefont
  {Denicol}}, \bibinfo {author} {\bibfnamefont {H.}~\bibnamefont {Niemi}},
  \bibinfo {author} {\bibfnamefont {E.}~\bibnamefont {Molnar}}, \ and\ \bibinfo
  {author} {\bibfnamefont {D.~H.}\ \bibnamefont {Rischke}},\ }\href {\doibase
  10.1103/PhysRevD.85.114047} {\bibfield  {journal} {\bibinfo  {journal} {Phys.
  Rev. D}\ }\textbf {\bibinfo {volume} {85}},\ \bibinfo {pages} {114047}
  (\bibinfo {year} {2012})}\BibitemShut {NoStop}%
\bibitem [{\citenamefont {Muronga}\ and\ \citenamefont
  {Rischke}(2004)}]{Muronga:2004sf}%
  \BibitemOpen
  \bibfield  {author} {\bibinfo {author} {\bibfnamefont {A.}~\bibnamefont
  {Muronga}}\ and\ \bibinfo {author} {\bibfnamefont {D.~H.}\ \bibnamefont
  {Rischke}},\ }\href@noop {} {\  (\bibinfo {year} {2004})},\ \Eprint
  {http://arxiv.org/abs/nucl-th/0407114} {arXiv:nucl-th/0407114 [nucl-th]}
  \BibitemShut {NoStop}%
%%CITATION = NUCL-TH/0407114;%%
\bibitem [{\citenamefont {Baier}\ \emph {et~al.}(2006)\citenamefont {Baier},
  \citenamefont {Romatschke},\ and\ \citenamefont {Wiedemann}}]{Baier:2006um}%
  \BibitemOpen
  \bibfield  {author} {\bibinfo {author} {\bibfnamefont {R.}~\bibnamefont
  {Baier}}, \bibinfo {author} {\bibfnamefont {P.}~\bibnamefont {Romatschke}}, \
  and\ \bibinfo {author} {\bibfnamefont {U.}~\bibnamefont {Wiedemann}},\ }\href
  {\doibase 10.1103/PhysRevC.73.064903} {\bibfield  {journal} {\bibinfo
  {journal} {Phys. Rev.}\ }\textbf {\bibinfo {volume} {C73}},\ \bibinfo {pages}
  {064903} (\bibinfo {year} {2006})},\ \Eprint
  {http://arxiv.org/abs/hep-ph/0602249} {arXiv:hep-ph/0602249 [hep-ph]}
  \BibitemShut {NoStop}%
%%CITATION = HEP-PH/0602249;%%
\bibitem [{\citenamefont {Song}\ and\ \citenamefont
  {Heinz}(2008{\natexlab{a}})}]{Song:2007fn}%
  \BibitemOpen
  \bibfield  {author} {\bibinfo {author} {\bibfnamefont {H.}~\bibnamefont
  {Song}}\ and\ \bibinfo {author} {\bibfnamefont {U.}~\bibnamefont {Heinz}},\
  }\href {\doibase 10.1016/j.physletb.2007.11.019} {\bibfield  {journal}
  {\bibinfo  {journal} {Phys. Lett.}\ }\textbf {\bibinfo {volume} {B658}},\
  \bibinfo {pages} {279} (\bibinfo {year} {2008}{\natexlab{a}})},\ \Eprint
  {http://arxiv.org/abs/0709.0742} {arXiv:0709.0742 [nucl-th]} \BibitemShut
  {NoStop}%
%%CITATION = ARXIV:0709.0742;%%
\bibitem [{\citenamefont {Song}\ and\ \citenamefont
  {Heinz}(2008{\natexlab{b}})}]{Song:2007ux}%
  \BibitemOpen
  \bibfield  {author} {\bibinfo {author} {\bibfnamefont {H.}~\bibnamefont
  {Song}}\ and\ \bibinfo {author} {\bibfnamefont {U.}~\bibnamefont {Heinz}},\
  }\href {\doibase 10.1103/PhysRevC.77.064901} {\bibfield  {journal} {\bibinfo
  {journal} {Phys. Rev.}\ }\textbf {\bibinfo {volume} {C77}},\ \bibinfo {pages}
  {064901} (\bibinfo {year} {2008}{\natexlab{b}})},\ \Eprint
  {http://arxiv.org/abs/0712.3715} {arXiv:0712.3715 [nucl-th]} \BibitemShut
  {NoStop}%
%%CITATION = ARXIV:0712.3715;%%
\bibitem [{\citenamefont {Muronga}(2002)}]{Muronga:2001zk}%
  \BibitemOpen
  \bibfield  {author} {\bibinfo {author} {\bibfnamefont {A.}~\bibnamefont
  {Muronga}},\ }\href {\doibase 10.1103/PhysRevLett.88.062302} {\bibfield
  {journal} {\bibinfo  {journal} {Phys. Rev. Lett.}\ }\textbf {\bibinfo
  {volume} {88}},\ \bibinfo {pages} {062302} (\bibinfo {year} {2002})},\
  \bibinfo {note} {[Erratum: Phys. Rev. Lett.89,159901(2002)]},\ \Eprint
  {http://arxiv.org/abs/nucl-th/0104064} {arXiv:nucl-th/0104064 [nucl-th]}
  \BibitemShut {NoStop}%
%%CITATION = NUCL-TH/0104064;%%
\bibitem [{\citenamefont {Muronga}(2004)}]{Muronga:2003ta}%
  \BibitemOpen
  \bibfield  {author} {\bibinfo {author} {\bibfnamefont {A.}~\bibnamefont
  {Muronga}},\ }\href {\doibase 10.1103/PhysRevC.69.034903} {\bibfield
  {journal} {\bibinfo  {journal} {Phys. Rev.}\ }\textbf {\bibinfo {volume}
  {C69}},\ \bibinfo {pages} {034903} (\bibinfo {year} {2004})},\ \Eprint
  {http://arxiv.org/abs/nucl-th/0309055} {arXiv:nucl-th/0309055 [nucl-th]}
  \BibitemShut {NoStop}%
%%CITATION = NUCL-TH/0309055;%%
\bibitem [{\citenamefont {Heinz}\ \emph {et~al.}(2006)\citenamefont {Heinz},
  \citenamefont {Song},\ and\ \citenamefont {Chaudhuri}}]{Heinz:2005bw}%
  \BibitemOpen
  \bibfield  {author} {\bibinfo {author} {\bibfnamefont {U.}~\bibnamefont
  {Heinz}}, \bibinfo {author} {\bibfnamefont {H.}~\bibnamefont {Song}}, \ and\
  \bibinfo {author} {\bibfnamefont {A.~K.}\ \bibnamefont {Chaudhuri}},\ }\href
  {\doibase 10.1103/PhysRevC.73.034904} {\bibfield  {journal} {\bibinfo
  {journal} {Phys.Rev.}\ }\textbf {\bibinfo {volume} {C73}},\ \bibinfo {pages}
  {034904} (\bibinfo {year} {2006})},\ \Eprint
  {http://arxiv.org/abs/nucl-th/0510014} {arXiv:nucl-th/0510014 [nucl-th]}
  \BibitemShut {NoStop}%
%%CITATION = NUCL-TH/0510014;%%
\bibitem [{\citenamefont {Muronga}(2007{\natexlab{a}})}]{Muronga:2006zx}%
  \BibitemOpen
  \bibfield  {author} {\bibinfo {author} {\bibfnamefont {A.}~\bibnamefont
  {Muronga}},\ }\href {\doibase 10.1103/PhysRevC.76.014910} {\bibfield
  {journal} {\bibinfo  {journal} {Phys. Rev.}\ }\textbf {\bibinfo {volume}
  {C76}},\ \bibinfo {pages} {014910} (\bibinfo {year} {2007}{\natexlab{a}})},\
  \Eprint {http://arxiv.org/abs/nucl-th/0611091} {arXiv:nucl-th/0611091
  [nucl-th]} \BibitemShut {NoStop}%
%%CITATION = NUCL-TH/0611091;%%
\bibitem [{\citenamefont {Muronga}(2007{\natexlab{b}})}]{Muronga:2006zw}%
  \BibitemOpen
  \bibfield  {author} {\bibinfo {author} {\bibfnamefont {A.}~\bibnamefont
  {Muronga}},\ }\href {\doibase 10.1103/PhysRevC.76.014909} {\bibfield
  {journal} {\bibinfo  {journal} {Phys. Rev.}\ }\textbf {\bibinfo {volume}
  {C76}},\ \bibinfo {pages} {014909} (\bibinfo {year} {2007}{\natexlab{b}})},\
  \Eprint {http://arxiv.org/abs/nucl-th/0611090} {arXiv:nucl-th/0611090
  [nucl-th]} \BibitemShut {NoStop}%
%%CITATION = NUCL-TH/0611090;%%
\bibitem [{\citenamefont {Romatschke}\ and\ \citenamefont
  {Strickland}(2004)}]{Romatschke:2003vc}%
  \BibitemOpen
  \bibfield  {author} {\bibinfo {author} {\bibfnamefont {P.}~\bibnamefont
  {Romatschke}}\ and\ \bibinfo {author} {\bibfnamefont {M.}~\bibnamefont
  {Strickland}},\ }\href {\doibase 10.1103/PhysRevD.69.065005} {\bibfield
  {journal} {\bibinfo  {journal} {Phys.Rev.}\ }\textbf {\bibinfo {volume}
  {D69}},\ \bibinfo {pages} {065005} (\bibinfo {year} {2004})},\ \Eprint
  {http://arxiv.org/abs/hep-ph/0309093} {arXiv:hep-ph/0309093 [hep-ph]}
  \BibitemShut {NoStop}%
%%CITATION = HEP-PH/0309093;%%
\bibitem [{\citenamefont {Molnar}\ \emph
  {et~al.}(2016{\natexlab{a}})\citenamefont {Molnar}, \citenamefont {Niemi},\
  and\ \citenamefont {Rischke}}]{Molnar:2016vvu}%
  \BibitemOpen
  \bibfield  {author} {\bibinfo {author} {\bibfnamefont {E.}~\bibnamefont
  {Molnar}}, \bibinfo {author} {\bibfnamefont {H.}~\bibnamefont {Niemi}}, \
  and\ \bibinfo {author} {\bibfnamefont {D.~H.}\ \bibnamefont {Rischke}},\
  }\href {\doibase 10.1103/PhysRevD.93.114025} {\bibfield  {journal} {\bibinfo
  {journal} {Phys. Rev.}\ }\textbf {\bibinfo {volume} {D93}},\ \bibinfo {pages}
  {114025} (\bibinfo {year} {2016}{\natexlab{a}})},\ \Eprint
  {http://arxiv.org/abs/1602.00573} {arXiv:1602.00573 [nucl-th]} \BibitemShut
  {NoStop}%
%%CITATION = ARXIV:1602.00573;%%
\bibitem [{\citenamefont {Molnar}\ \emph
  {et~al.}(2016{\natexlab{b}})\citenamefont {Molnar}, \citenamefont {Niemi},\
  and\ \citenamefont {Rischke}}]{Molnar:2016gwq}%
  \BibitemOpen
  \bibfield  {author} {\bibinfo {author} {\bibfnamefont {E.}~\bibnamefont
  {Molnar}}, \bibinfo {author} {\bibfnamefont {H.}~\bibnamefont {Niemi}}, \
  and\ \bibinfo {author} {\bibfnamefont {D.~H.}\ \bibnamefont {Rischke}},\
  }\href {\doibase 10.1103/PhysRevD.94.125003} {\bibfield  {journal} {\bibinfo
  {journal} {Phys. Rev.}\ }\textbf {\bibinfo {volume} {D94}},\ \bibinfo {pages}
  {125003} (\bibinfo {year} {2016}{\natexlab{b}})},\ \Eprint
  {http://arxiv.org/abs/1606.09019} {arXiv:1606.09019 [nucl-th]} \BibitemShut
  {NoStop}%
%%CITATION = ARXIV:1606.09019;%%
\bibitem [{\citenamefont {Tinti}(2015)}]{Tinti:2015xwa}%
  \BibitemOpen
  \bibfield  {author} {\bibinfo {author} {\bibfnamefont {L.}~\bibnamefont
  {Tinti}},\ }\href@noop {} {\  (\bibinfo {year} {2015})},\ \Eprint
  {http://arxiv.org/abs/1506.07164} {arXiv:1506.07164 [hep-ph]} \BibitemShut
  {NoStop}%
%%CITATION = ARXIV:1506.07164;%%
\bibitem [{\citenamefont {Baym}(1984)}]{Baym:1984np}%
  \BibitemOpen
  \bibfield  {author} {\bibinfo {author} {\bibfnamefont {G.}~\bibnamefont
  {Baym}},\ }\href@noop {} {\bibfield  {journal} {\bibinfo  {journal} {Phys.
  Lett.}\ }\textbf {\bibinfo {volume} {B138}},\ \bibinfo {pages} {18} (\bibinfo
  {year} {1984})}\BibitemShut {NoStop}%
%%CITATION = PHLTA,B138,18;%%
\bibitem [{\citenamefont {Florkowski}\ \emph
  {et~al.}(2013{\natexlab{b}})\citenamefont {Florkowski}, \citenamefont
  {Ryblewski},\ and\ \citenamefont {Strickland}}]{Florkowski:2013lza}%
  \BibitemOpen
  \bibfield  {author} {\bibinfo {author} {\bibfnamefont {W.}~\bibnamefont
  {Florkowski}}, \bibinfo {author} {\bibfnamefont {R.}~\bibnamefont
  {Ryblewski}}, \ and\ \bibinfo {author} {\bibfnamefont {M.}~\bibnamefont
  {Strickland}},\ }\href {\doibase 10.1016/j.nuclphysa.2013.08.004} {\bibfield
  {journal} {\bibinfo  {journal} {Nucl.Phys.}\ }\textbf {\bibinfo {volume}
  {A916}},\ \bibinfo {pages} {249} (\bibinfo {year} {2013}{\natexlab{b}})},\
  \Eprint {http://arxiv.org/abs/1304.0665} {arXiv:1304.0665 [nucl-th]}
  \BibitemShut {NoStop}%
%%CITATION = ARXIV:1304.0665;%%
\bibitem [{\citenamefont {Florkowski}\ \emph
  {et~al.}(2013{\natexlab{c}})\citenamefont {Florkowski}, \citenamefont
  {Ryblewski},\ and\ \citenamefont {Strickland}}]{Florkowski:2013lya}%
  \BibitemOpen
  \bibfield  {author} {\bibinfo {author} {\bibfnamefont {W.}~\bibnamefont
  {Florkowski}}, \bibinfo {author} {\bibfnamefont {R.}~\bibnamefont
  {Ryblewski}}, \ and\ \bibinfo {author} {\bibfnamefont {M.}~\bibnamefont
  {Strickland}},\ }\href {\doibase 10.1103/PhysRevC.88.024903} {\bibfield
  {journal} {\bibinfo  {journal} {Phys. Rev.}\ }\textbf {\bibinfo {volume}
  {C88}},\ \bibinfo {pages} {024903} (\bibinfo {year} {2013}{\natexlab{c}})},\
  \Eprint {http://arxiv.org/abs/1305.7234} {arXiv:1305.7234 [nucl-th]}
  \BibitemShut {NoStop}%
%%CITATION = ARXIV:1305.7234;%%
\bibitem [{\citenamefont {Florkowski}\ \emph
  {et~al.}(2014{\natexlab{b}})\citenamefont {Florkowski}, \citenamefont
  {Maksymiuk}, \citenamefont {Ryblewski},\ and\ \citenamefont
  {Strickland}}]{Florkowski:2014sfa}%
  \BibitemOpen
  \bibfield  {author} {\bibinfo {author} {\bibfnamefont {W.}~\bibnamefont
  {Florkowski}}, \bibinfo {author} {\bibfnamefont {E.}~\bibnamefont
  {Maksymiuk}}, \bibinfo {author} {\bibfnamefont {R.}~\bibnamefont
  {Ryblewski}}, \ and\ \bibinfo {author} {\bibfnamefont {M.}~\bibnamefont
  {Strickland}},\ }\href@noop {} {\  (\bibinfo {year} {2014}{\natexlab{b}})},\
  \Eprint {http://arxiv.org/abs/1402.7348} {arXiv:1402.7348 [hep-ph]}
  \BibitemShut {NoStop}%
%%CITATION = ARXIV:1402.7348;%%
\bibitem [{\citenamefont {Denicol}\ \emph
  {et~al.}(2014{\natexlab{b}})\citenamefont {Denicol}, \citenamefont {Heinz},
  \citenamefont {Martinez}, \citenamefont {Noronha},\ and\ \citenamefont
  {Strickland}}]{Denicol:2014tha}%
  \BibitemOpen
  \bibfield  {author} {\bibinfo {author} {\bibfnamefont {G.~S.}\ \bibnamefont
  {Denicol}}, \bibinfo {author} {\bibfnamefont {U.}~\bibnamefont {Heinz}},
  \bibinfo {author} {\bibfnamefont {M.}~\bibnamefont {Martinez}}, \bibinfo
  {author} {\bibfnamefont {J.}~\bibnamefont {Noronha}}, \ and\ \bibinfo
  {author} {\bibfnamefont {M.}~\bibnamefont {Strickland}},\ }\href {\doibase
  10.1103/PhysRevD.90.125026} {\bibfield  {journal} {\bibinfo  {journal} {Phys.
  Rev. D}\ }\textbf {\bibinfo {volume} {90}},\ \bibinfo {pages} {125026}
  (\bibinfo {year} {2014}{\natexlab{b}})},\ \Eprint
  {http://arxiv.org/abs/1408.7048} {arXiv:1408.7048 [hep-ph]} \BibitemShut
  {NoStop}%
%%CITATION = ARXIV:1408.7048;%%
\bibitem [{\citenamefont {Denicol}\ \emph
  {et~al.}(2014{\natexlab{c}})\citenamefont {Denicol}, \citenamefont {Heinz},
  \citenamefont {Martinez}, \citenamefont {Noronha},\ and\ \citenamefont
  {Strickland}}]{Denicol:2014xca}%
  \BibitemOpen
  \bibfield  {author} {\bibinfo {author} {\bibfnamefont {G.~S.}\ \bibnamefont
  {Denicol}}, \bibinfo {author} {\bibfnamefont {U.}~\bibnamefont {Heinz}},
  \bibinfo {author} {\bibfnamefont {M.}~\bibnamefont {Martinez}}, \bibinfo
  {author} {\bibfnamefont {J.}~\bibnamefont {Noronha}}, \ and\ \bibinfo
  {author} {\bibfnamefont {M.}~\bibnamefont {Strickland}},\ }\href {\doibase
  10.1103/PhysRevLett.113.202301} {\bibfield  {journal} {\bibinfo  {journal}
  {Phys. Rev. Lett.}\ }\textbf {\bibinfo {volume} {113}},\ \bibinfo {pages}
  {202301} (\bibinfo {year} {2014}{\natexlab{c}})},\ \Eprint
  {http://arxiv.org/abs/1408.5646} {arXiv:1408.5646 [hep-ph]} \BibitemShut
  {NoStop}%
%%CITATION = ARXIV:1408.5646;%%
\bibitem [{\citenamefont {Heinz}\ and\ \citenamefont
  {Martinez}(2015)}]{Heinz:2015cda}%
  \BibitemOpen
  \bibfield  {author} {\bibinfo {author} {\bibfnamefont {U.}~\bibnamefont
  {Heinz}}\ and\ \bibinfo {author} {\bibfnamefont {M.}~\bibnamefont
  {Martinez}},\ }\href {\doibase 10.1016/j.nuclphysa.2015.08.009} {\bibfield
  {journal} {\bibinfo  {journal} {Nucl. Phys.}\ }\textbf {\bibinfo {volume}
  {A943}},\ \bibinfo {pages} {26} (\bibinfo {year} {2015})},\ \Eprint
  {http://arxiv.org/abs/1506.07500} {arXiv:1506.07500 [hep-ph]} \BibitemShut
  {NoStop}%
%%CITATION = ARXIV:1506.07500;%%
\bibitem [{\citenamefont {Gubser}(2010)}]{Gubser:2010ze}%
  \BibitemOpen
  \bibfield  {author} {\bibinfo {author} {\bibfnamefont {S.~S.}\ \bibnamefont
  {Gubser}},\ }\href {\doibase 10.1103/PhysRevD.82.085027} {\bibfield
  {journal} {\bibinfo  {journal} {Phys. Rev.}\ }\textbf {\bibinfo {volume}
  {D82}},\ \bibinfo {pages} {085027} (\bibinfo {year} {2010})},\ \Eprint
  {http://arxiv.org/abs/1006.0006} {arXiv:1006.0006 [hep-th]} \BibitemShut
  {NoStop}%
%%CITATION = ARXIV:1006.0006;%%
\bibitem [{\citenamefont {Gubser}\ and\ \citenamefont
  {Yarom}(2011)}]{Gubser:2010ui}%
  \BibitemOpen
  \bibfield  {author} {\bibinfo {author} {\bibfnamefont {S.~S.}\ \bibnamefont
  {Gubser}}\ and\ \bibinfo {author} {\bibfnamefont {A.}~\bibnamefont {Yarom}},\
  }\href {\doibase 10.1016/j.nuclphysb.2011.01.012} {\bibfield  {journal}
  {\bibinfo  {journal} {Nucl. Phys.}\ }\textbf {\bibinfo {volume} {B846}},\
  \bibinfo {pages} {469} (\bibinfo {year} {2011})},\ \Eprint
  {http://arxiv.org/abs/1012.1314} {arXiv:1012.1314 [hep-th]} \BibitemShut
  {NoStop}%
%%CITATION = ARXIV:1012.1314;%%
\bibitem [{\citenamefont {Nopoush}\ \emph
  {et~al.}(2015{\natexlab{b}})\citenamefont {Nopoush}, \citenamefont
  {Ryblewski},\ and\ \citenamefont {Strickland}}]{Nopoush:2014qba}%
  \BibitemOpen
  \bibfield  {author} {\bibinfo {author} {\bibfnamefont {M.}~\bibnamefont
  {Nopoush}}, \bibinfo {author} {\bibfnamefont {R.}~\bibnamefont {Ryblewski}},
  \ and\ \bibinfo {author} {\bibfnamefont {M.}~\bibnamefont {Strickland}},\
  }\href {\doibase 10.1103/PhysRevD.91.045007} {\bibfield  {journal} {\bibinfo
  {journal} {Phys. Rev. D}\ }\textbf {\bibinfo {volume} {91}},\ \bibinfo
  {pages} {045007} (\bibinfo {year} {2015}{\natexlab{b}})},\ \Eprint
  {http://arxiv.org/abs/1410.6790} {arXiv:1410.6790 [nucl-th]} \BibitemShut
  {NoStop}%
%%CITATION = ARXIV:1410.6790;%%
\bibitem [{\citenamefont {Bazow}\ \emph
  {et~al.}(2016{\natexlab{b}})\citenamefont {Bazow}, \citenamefont {Denicol},
  \citenamefont {Heinz}, \citenamefont {Martinez},\ and\ \citenamefont
  {Noronha}}]{Bazow:2016oky}%
  \BibitemOpen
  \bibfield  {author} {\bibinfo {author} {\bibfnamefont {D.}~\bibnamefont
  {Bazow}}, \bibinfo {author} {\bibfnamefont {G.~S.}\ \bibnamefont {Denicol}},
  \bibinfo {author} {\bibfnamefont {U.}~\bibnamefont {Heinz}}, \bibinfo
  {author} {\bibfnamefont {M.}~\bibnamefont {Martinez}}, \ and\ \bibinfo
  {author} {\bibfnamefont {J.}~\bibnamefont {Noronha}},\ }\href {\doibase
  10.1103/PhysRevD.94.125006} {\bibfield  {journal} {\bibinfo  {journal} {Phys.
  Rev.}\ }\textbf {\bibinfo {volume} {D94}},\ \bibinfo {pages} {125006}
  (\bibinfo {year} {2016}{\natexlab{b}})},\ \Eprint
  {http://arxiv.org/abs/1607.05245} {arXiv:1607.05245 [hep-ph]} \BibitemShut
  {NoStop}%
%%CITATION = ARXIV:1607.05245;%%
\bibitem [{\citenamefont {Landau}\ and\ \citenamefont
  {Lifshitz}(1987)}]{Landaufluid}%
  \BibitemOpen
  \bibfield  {author} {\bibinfo {author} {\bibfnamefont {L.~D.}\ \bibnamefont
  {Landau}}\ and\ \bibinfo {author} {\bibfnamefont {E.~M.}\ \bibnamefont
  {Lifshitz}},\ }\href@noop {} {\emph {\bibinfo {title} {Fluid Mechanics,
  Second Edition: Volume 6 (Course of Theoretical Physics)}}}\ (\bibinfo
  {publisher} {Butterworth-Heinemann},\ \bibinfo {year} {1987})\BibitemShut
  {NoStop}%
\bibitem [{\citenamefont {Denicol}(2014)}]{Denicol:2014loa}%
  \BibitemOpen
  \bibfield  {author} {\bibinfo {author} {\bibfnamefont {G.~S.}\ \bibnamefont
  {Denicol}},\ }\href {\doibase 10.1088/0954-3899/41/12/124004} {\bibfield
  {journal} {\bibinfo  {journal} {J. Phys.}\ }\textbf {\bibinfo {volume}
  {G41}},\ \bibinfo {pages} {124004} (\bibinfo {year} {2014})}\BibitemShut
  {NoStop}%
%%CITATION = JPAGA,G41,124004;%%
\bibitem [{\citenamefont {Huang}\ \emph {et~al.}(2010)\citenamefont {Huang},
  \citenamefont {Huang}, \citenamefont {Rischke},\ and\ \citenamefont
  {Sedrakian}}]{Huang:2009ue}%
  \BibitemOpen
  \bibfield  {author} {\bibinfo {author} {\bibfnamefont {X.-G.}\ \bibnamefont
  {Huang}}, \bibinfo {author} {\bibfnamefont {M.}~\bibnamefont {Huang}},
  \bibinfo {author} {\bibfnamefont {D.~H.}\ \bibnamefont {Rischke}}, \ and\
  \bibinfo {author} {\bibfnamefont {A.}~\bibnamefont {Sedrakian}},\ }\href
  {\doibase 10.1103/PhysRevD.81.045015} {\bibfield  {journal} {\bibinfo
  {journal} {Phys. Rev.}\ }\textbf {\bibinfo {volume} {D81}},\ \bibinfo {pages}
  {045015} (\bibinfo {year} {2010})},\ \Eprint {http://arxiv.org/abs/0910.3633}
  {arXiv:0910.3633 [astro-ph.HE]} \BibitemShut {NoStop}%
%%CITATION = ARXIV:0910.3633;%%
\bibitem [{\citenamefont {Huang}\ \emph {et~al.}(2011)\citenamefont {Huang},
  \citenamefont {Sedrakian},\ and\ \citenamefont {Rischke}}]{Huang:2011dc}%
  \BibitemOpen
  \bibfield  {author} {\bibinfo {author} {\bibfnamefont {X.-G.}\ \bibnamefont
  {Huang}}, \bibinfo {author} {\bibfnamefont {A.}~\bibnamefont {Sedrakian}}, \
  and\ \bibinfo {author} {\bibfnamefont {D.~H.}\ \bibnamefont {Rischke}},\
  }\href {\doibase 10.1016/j.aop.2011.08.001} {\bibfield  {journal} {\bibinfo
  {journal} {Annals Phys.}\ }\textbf {\bibinfo {volume} {326}},\ \bibinfo
  {pages} {3075} (\bibinfo {year} {2011})},\ \Eprint
  {http://arxiv.org/abs/1108.0602} {arXiv:1108.0602 [astro-ph.HE]} \BibitemShut
  {NoStop}%
%%CITATION = ARXIV:1108.0602;%%
\bibitem [{\citenamefont {Gedalin}(1991)}]{Gedalin1}%
  \BibitemOpen
  \bibfield  {author} {\bibinfo {author} {\bibfnamefont {M.}~\bibnamefont
  {Gedalin}},\ }\href {\doibase http://dx.doi.org/10.1063/1.859656} {\bibfield
  {journal} {\bibinfo  {journal} {Physics of Fluids B}\ }\textbf {\bibinfo
  {volume} {3}},\ \bibinfo {pages} {1871} (\bibinfo {year} {1991})}\BibitemShut
  {NoStop}%
\bibitem [{\citenamefont {Gedalin}\ and\ \citenamefont
  {Oiberman}(1995)}]{Gedalin2}%
  \BibitemOpen
  \bibfield  {author} {\bibinfo {author} {\bibfnamefont {M.}~\bibnamefont
  {Gedalin}}\ and\ \bibinfo {author} {\bibfnamefont {I.}~\bibnamefont
  {Oiberman}},\ }\href {\doibase 10.1103/PhysRevE.51.4901} {\bibfield
  {journal} {\bibinfo  {journal} {Phys. Rev. E}\ }\textbf {\bibinfo {volume}
  {51}},\ \bibinfo {pages} {4901} (\bibinfo {year} {1995})}\BibitemShut
  {NoStop}%
\bibitem [{\citenamefont {Romatschke}\ and\ \citenamefont
  {Strickland}(2003)}]{Romatschke:2003ms}%
  \BibitemOpen
  \bibfield  {author} {\bibinfo {author} {\bibfnamefont {P.}~\bibnamefont
  {Romatschke}}\ and\ \bibinfo {author} {\bibfnamefont {M.}~\bibnamefont
  {Strickland}},\ }\href@noop {} {\bibfield  {journal} {\bibinfo  {journal}
  {Phys. Rev.}\ }\textbf {\bibinfo {volume} {D68}},\ \bibinfo {pages} {036004}
  (\bibinfo {year} {2003})},\ \Eprint {http://arxiv.org/abs/hep-ph/0304092}
  {hep-ph/0304092} \BibitemShut {NoStop}%
%%CITATION = HEP-PH 0304092;%%
\bibitem [{\citenamefont {Banerjee}\ \emph {et~al.}(1989)\citenamefont
  {Banerjee}, \citenamefont {Bhalerao},\ and\ \citenamefont
  {Ravishankar}}]{Banerjee:1989by}%
  \BibitemOpen
  \bibfield  {author} {\bibinfo {author} {\bibfnamefont {B.}~\bibnamefont
  {Banerjee}}, \bibinfo {author} {\bibfnamefont {R.~S.}\ \bibnamefont
  {Bhalerao}}, \ and\ \bibinfo {author} {\bibfnamefont {V.}~\bibnamefont
  {Ravishankar}},\ }\href {\doibase 10.1016/0370-2693(89)91041-1} {\bibfield
  {journal} {\bibinfo  {journal} {Phys. Lett.}\ }\textbf {\bibinfo {volume}
  {B224}},\ \bibinfo {pages} {16} (\bibinfo {year} {1989})}\BibitemShut
  {NoStop}%
%%CITATION = PHLTA,B224,16;%%
\bibitem [{\citenamefont {Policastro}\ \emph {et~al.}(2001)\citenamefont
  {Policastro}, \citenamefont {Son},\ and\ \citenamefont
  {Starinets}}]{Policastro:2001yc}%
  \BibitemOpen
  \bibfield  {author} {\bibinfo {author} {\bibfnamefont {G.}~\bibnamefont
  {Policastro}}, \bibinfo {author} {\bibfnamefont {D.~T.}\ \bibnamefont {Son}},
  \ and\ \bibinfo {author} {\bibfnamefont {A.~O.}\ \bibnamefont {Starinets}},\
  }\href {\doibase 10.1103/PhysRevLett.87.081601} {\bibfield  {journal}
  {\bibinfo  {journal} {Phys. Rev. Lett.}\ }\textbf {\bibinfo {volume} {87}},\
  \bibinfo {pages} {081601} (\bibinfo {year} {2001})},\ \Eprint
  {http://arxiv.org/abs/hep-th/0104066} {arXiv:hep-th/0104066} \BibitemShut
  {NoStop}%
%%CITATION = HEP-TH/0104066;%%
\bibitem [{\citenamefont {Kovtun}\ \emph {et~al.}(2005)\citenamefont {Kovtun},
  \citenamefont {Son},\ and\ \citenamefont {Starinets}}]{Kovtun:2004de}%
  \BibitemOpen
  \bibfield  {author} {\bibinfo {author} {\bibfnamefont {P.}~\bibnamefont
  {Kovtun}}, \bibinfo {author} {\bibfnamefont {D.}~\bibnamefont {Son}}, \ and\
  \bibinfo {author} {\bibfnamefont {A.}~\bibnamefont {Starinets}},\ }\href
  {\doibase 10.1103/PhysRevLett.94.111601} {\bibfield  {journal} {\bibinfo
  {journal} {Phys. Rev. Lett.}\ }\textbf {\bibinfo {volume} {94}},\ \bibinfo
  {pages} {111601} (\bibinfo {year} {2005})},\ \Eprint
  {http://arxiv.org/abs/hep-th/0405231} {arXiv:hep-th/0405231 [hep-th]}
  \BibitemShut {NoStop}%
%%CITATION = HEP-TH/0405231;%%
\bibitem [{\citenamefont {Romatschke}(2017)}]{Romatschke:2016hle}%
  \BibitemOpen
  \bibfield  {author} {\bibinfo {author} {\bibfnamefont {P.}~\bibnamefont
  {Romatschke}},\ }\href {\doibase 10.1140/epjc/s10052-016-4567-x} {\bibfield
  {journal} {\bibinfo  {journal} {Eur. Phys. J.}\ }\textbf {\bibinfo {volume}
  {C77}},\ \bibinfo {pages} {21} (\bibinfo {year} {2017})},\ \Eprint
  {http://arxiv.org/abs/1609.02820} {arXiv:1609.02820 [nucl-th]} \BibitemShut
  {NoStop}%
%%CITATION = ARXIV:1609.02820;%%
\bibitem [{\citenamefont {Israel}\ and\ \citenamefont
  {Stewart}(1979)}]{Israel:1979wp}%
  \BibitemOpen
  \bibfield  {author} {\bibinfo {author} {\bibfnamefont {W.}~\bibnamefont
  {Israel}}\ and\ \bibinfo {author} {\bibfnamefont {J.~M.}\ \bibnamefont
  {Stewart}},\ }\href {\doibase 10.1016/0003-4916(79)90130-1} {\bibfield
  {journal} {\bibinfo  {journal} {Ann. Phys.}\ }\textbf {\bibinfo {volume}
  {118}},\ \bibinfo {pages} {341} (\bibinfo {year} {1979})}\BibitemShut
  {NoStop}%
%%CITATION = APNYA,118,341;%%
\bibitem [{\citenamefont {Baier}\ \emph {et~al.}(2008)\citenamefont {Baier},
  \citenamefont {Romatschke}, \citenamefont {Son}, \citenamefont {Starinets},\
  and\ \citenamefont {Stephanov}}]{Baier:2007ix}%
  \BibitemOpen
  \bibfield  {author} {\bibinfo {author} {\bibfnamefont {R.}~\bibnamefont
  {Baier}}, \bibinfo {author} {\bibfnamefont {P.}~\bibnamefont {Romatschke}},
  \bibinfo {author} {\bibfnamefont {D.~T.}\ \bibnamefont {Son}}, \bibinfo
  {author} {\bibfnamefont {A.~O.}\ \bibnamefont {Starinets}}, \ and\ \bibinfo
  {author} {\bibfnamefont {M.~A.}\ \bibnamefont {Stephanov}},\ }\href {\doibase
  10.1088/1126-6708/2008/04/100} {\bibfield  {journal} {\bibinfo  {journal}
  {JHEP}\ }\textbf {\bibinfo {volume} {0804}},\ \bibinfo {pages} {100}
  (\bibinfo {year} {2008})},\ \Eprint {http://arxiv.org/abs/0712.2451}
  {arXiv:0712.2451 [hep-th]} \BibitemShut {NoStop}%
%%CITATION = 0712.2451;%%
\end{thebibliography}%

\end{document}